%% file: main.tex
\documentclass[aps,prl,reprint,superscriptaddress,amsmath,amssymb,floatfix]{revtex4-2}
\usepackage{graphicx}
\usepackage{xcolor}

\usepackage[hidelinks]{hyperref}

\newlength{\resultcolumnwidth}

\begin{document}
\title{Accelerating Quantum Error Correction Simulations under Coherent Noise with Compiled Clifford-Augmented Matrix Product States}
\author{Jiale Huang}
\affiliation{Key Laboratory of Artificial Structures and Quantum Control (Ministry of Education), School of Physics and Astronomy, Shanghai Jiao Tong University, Shanghai 200240, China}
\affiliation{FieldQuantum Research Institute, Beijing 100176, China}
\affiliation{FieldQuantum Frontier Research Institute, Beijing 100190, China}

\author{Yaojian Chen}
\affiliation{FieldQuantum Research Institute, Beijing 100176, China}
\affiliation{FieldQuantum Frontier Research Institute, Beijing 100190, China}

\author{Ke Liu}
\affiliation{Hefei National Research Center for Physical Sciences at the Microscale and School of Physical Sciences, University of Science and Technology of China, Hefei 230026, China}
\affiliation{Shanghai Research Center for Quantum Science and CAS Center for Excellence in Quantum Information and Quantum Physics, University of Science and Technology of China, Shanghai 201315, China}

\author{Dingshun Lv}
\email[Corresponding author: ]{lvdingshun@fieldquanta.com}
\affiliation{FieldQuantum Research Institute, Beijing 100176, China}
\affiliation{FieldQuantum Frontier Research Institute, Beijing 100190, China}

\author{Mingpu Qin}
\email[Corresponding author: ]{qinmingpu@sjtu.edu.cn}
\affiliation{Key Laboratory of Artificial Structures and Quantum Control (Ministry of Education), School of Physics and Astronomy, Shanghai Jiao Tong University, Shanghai 200240, China}
\begin{abstract}
Reliable assessment of quantum error correction requires understanding how coherent errors affect logical performance.
Although tensor networks naturally capture coherent interference, entanglement growth and large qubit counts make repeated sampling costly. Here we develop a compiled Clifford-Augmented Matrix Product State (CAMPS) framework that reuses Clifford transformations across shots, recycles qubits released by measurements, evolves a compressed residual state, and accelerates sampling through rotation fusion and compilation tailored to the MPS layout. Surface-code benchmarks under fixed-sign coherent noise yield accurate logical error estimates at small bond dimensions, while random-sign simulations up to distance $d=9$ and on the $[[72,12,6]]$ bivariate bicycle code match the Pauli-twirling approximation results. Compared with uncompiled CAMPS, compiled CAMPS converges at smaller bond dimensions and achieves order-of-magnitude sampling speedups at equal bond dimension, opening a scalable route to studying coherent error propagation in larger fault-tolerant circuits.
\end{abstract}
\maketitle
\setlength{\resultcolumnwidth}{\dimexpr(\textwidth-\columnsep)/2\relax}

\begin{figure*}[t]
\centering
\includegraphics[width=\textwidth]{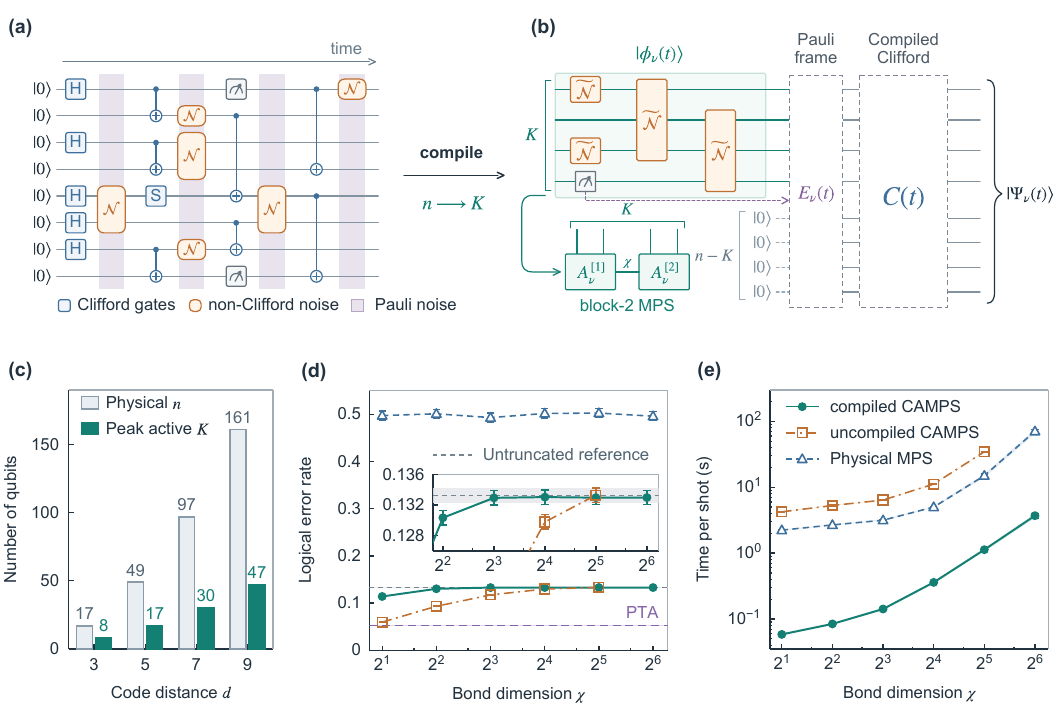}
\caption{Compiled CAMPS workflow and surface-code benchmarks.
(a,b) Orange blocks show physical noise $\mathcal N$ and its compiled residual action $\widetilde{\mathcal N}$. The compiled $C(t)$ is shared by all shots; $E_\nu(t)$ and the block-2 MPS tensors $A_\nu^{[a]}$ vary between shots. The MPS uses $K=\max_t k(t)$ positions.
(c) Physical-qubit count $n$ and peak active-qubit count $K$ versus code distance $d$ for surface-code $X$ memory under coherent $ZZ$ noise, with $r=d$ rounds.
(d,e) Fixed-sign $ZZ$ benchmark at $d=r=5$, $\theta=0.02$, and $p=0.005$.
(d) Logical error rate $p_L$ versus bond dimension $\chi$. Gray and purple dashed lines mark the untruncated $K=17$ reference and PTA prediction. Compiled and uncompiled CAMPS and the state-vector reference each use $5\times10^5$ shots. Bars and shading show 95\% confidence intervals. Inset: zoom near the untruncated reference.
(e) Median sampling time per shot versus $\chi$. Compiled CAMPS is more than an order of magnitude faster than uncompiled CAMPS and physical MPS at equal $\chi$.}
\label{fig:camps}
\end{figure*}

\emph{Introduction.---}Predicting how physical noise translates into logical failure under quantum error correction (QEC) is essential to building fault-tolerant quantum computers.
Coherent errors complicate this task because amplitudes for distinct error paths can interfere, an effect discarded by stochastic Pauli models~\cite{bravyi_coherent,huang}.
Tensor-network methods~\cite{RevModPhys.93.045003,xiang2023density}, particularly matrix product states (MPS), provide a natural framework for calculate these amplitudes and their interference~\cite{white1992,ostlund1995,vidal,schollwock}.
As entanglement accumulates, however, the bond dimensions required for an accurate description can make repeated trajectory sampling prohibitively expensive.

In earlier work, we introduced Clifford-Augmented Matrix Product States (CAMPS) for simulating quantum many-body systems~\cite{camps}, representing a state as an MPS augmented by a Clifford circuit. Optimizing this Clifford representation reduces the residual MPS entanglement while preserving the physical state.
In Hamiltonian simulations of ground states~\cite{critical,SciPostPhys.18.5.165,fermion,kitaev,grassmann,matchgates,10.1021/acs.jctc.6c01170}, finite-temperature properties~\cite{thermal}, and real-time dynamics~\cite{tdvp,dressedtdvp,fluxtube,aklt}, disentanglers are updated iteratively across MPS bonds while optimizing or evolving a single state.

Quantum circuit simulation differs from Hamiltonian evolution in both structure and sampling cost.
Instead of acting through coupled Hamiltonian terms, fault-tolerant circuits interleave explicit Clifford gates with sparse non-Clifford gates or coherent errors, leaving the residual state weakly entangled even at non-Clifford gate counts comparable to system size~\cite{nsee,ofd,fux,magicstn}.
Estimating logical error rates also requires sampling many noisy trajectories of the same circuit rather than evolving a single state, motivating recent hybrid simulators~\cite{mpstab} and MPS/CAMPS studies of coherent surface-code noise~\cite{harper}, QEC circuits, and magic-state cultivation~\cite{optimizedmps,cultivation}.

In such circuits, mid-circuit QEC measurements leave subsets of qubits isolated from the rest of the state.
Using these free qubits, suitable Clifford transformations can turn subsequent multi-qubit Pauli rotations or projections into local single-qubit updates without increasing the MPS bond dimension~\cite{iccr,ofd}.
Recent compilers build on this property, with Clifft compiling qubit recycling and Clifford evolution from circuit structure~\cite{clifft} and SymFT encoding fault and feedback dependence symbolically across trajectories~\cite{symft}.

Although Clifford compilation and CAMPS exploit the same representation freedom, existing compilers fix a Clifford sequence without tailoring the residual basis for MPS compression, whereas dynamic CAMPS updates the basis separately in each shot without recycling qubits.
Capturing both advantages requires coordinating qubit recycling and residual-basis selection entirely at compile time from circuit structure alone.

In this Letter, we develop compiled CAMPS, combining cross-shot Clifford compilation and free-qubit recycling with block-MPS evolution, rotation fusion, and operator-span minimization.
On rotated surface codes up to $d=9$~\cite{bombin2007,fowler2012} and the $[[72,12,6]]$ bivariate bicycle (BB) code~\cite{bb} under coherent noise, compiled CAMPS converges at smaller bond dimensions $\chi$ and samples over an order of magnitude faster than uncompiled CAMPS at equal $\chi$.

\emph{Compiled CAMPS.---}CAMPS represents a quantum state as $\lvert\Psi\rangle=C\lvert\psi[A]\rangle$, with a Clifford operator $C$ augmenting an MPS specified by tensors $A$~\cite{camps}.
Stochastic Pauli errors and measurement corrections are collected into an operator called the Pauli frame.
Before sampling, the compiler constructs a shared Clifford sequence $\{C(t)\}$ and a symbolic frame $E(t;\mathbf b)$, where $\mathbf b$ records noise realizations and measurement outcomes~\cite{clifft,symft}.
At circuit step $t$, shot $\nu$ evaluates the frame on its sampled record $\mathbf b_\nu$,
\begin{equation}
 \begin{aligned}
 \lvert\Psi_\nu(t)\rangle
 &=C(t)E_\nu(t)\lvert\psi[A_\nu(t)]\rangle,\\
 E(t;\mathbf b)&\xrightarrow[\mathbf b=\mathbf b_\nu]{\mathrm{shot}\,\nu}E_\nu(t).
 \end{aligned}
 \label{eq:compiled-state}
\end{equation}
The residual state $\lvert\psi[A_\nu(t)]\rangle$ contains $k(t)$ active qubits and free qubits in known $\lvert0\rangle$ state.
The compiler tracks their activation and release, reserving $K=\max_t k(t)$ MPS positions.
For $n$ physical qubits, pairing neighboring positions into $L=\lceil K/2\rceil$ blocks and omitting the step label $t$ hereafter gives
\begin{equation}
 \begin{aligned}
 \lvert\psi[A_\nu]\rangle
 &=\lvert\phi_\nu\rangle\otimes\lvert0\rangle^{\otimes(n-K)},\\
 \lvert\phi_\nu\rangle
 &=\sum_{\{s_a\}} A_\nu^{[1]s_1}\cdots A_\nu^{[L]s_L}
 \lvert s_1\cdots s_L\rangle .
 \end{aligned}
 \label{eq:block-state}
\end{equation}
Each block index $s_a$ has dimension four (or two for an unpaired final qubit), inter-block bonds have dimensions $\chi_a\leq\chi$, and the $K-k(t)$ vacant positions remain in $\lvert0\rangle$. The effect of block size is analyzed in the End Matter (Fig.~\ref{fig:block-mps}).
In each shot, the residual tensors $A_\nu$ evolve under compiled non-Clifford rotations and measurements, as illustrated in Fig.~\ref{fig:camps}(a,b).

We now construct the shared Clifford sequence and symbolic frame $E(\mathbf b)\equiv E(t;\mathbf b)$, with ``compile'' arrows denoting updates made before sampling.
A physical Clifford gate $U$ gives $C\xrightarrow{\rm compile}UC$, whereas a Pauli error $F^e$ ($e=0$ for $I$ and $e=1$ for $F$) satisfies $F^eC=C(C^\dagger FC)^e$ and updates \mbox{$E(\mathbf b)\xrightarrow{\rm compile}(C^\dagger FC)^eE(\mathbf b)$}, keeping $e$ symbolic until sampled in each shot.
A non-Clifford noise rotation $\mathcal N_j=e^{i\theta_jP_j}$ then acts on the residual state as
\begin{equation}
 \begin{aligned}
 \widetilde{\mathcal N}_{j,\nu}
 &=E_\nu^\dagger C^\dagger\mathcal N_jCE_\nu\\
 &=e^{i\vartheta_{j,\nu}Q_j},\qquad
 \vartheta_{j,\nu}=(-1)^{\ell_j(\mathbf b_\nu)}\theta_j .
 \end{aligned}
 \label{eq:compiled-rotation}
\end{equation}
Here $Q_j$ is the unsigned Pauli string in $C^\dagger P_jC=\pm Q_j$, while the binary function $\ell_j(\mathbf b)$ tracks this sign together with sign flips from anticommuting Pauli errors and measurement corrections.
Both $Q_j$ and $\ell_j$ are fixed during compilation, and each shot evaluates $\ell_j(\mathbf b_\nu)$ to determine its effective rotation angle~\cite{symft}.

Rotations sharing the same Pauli string $Q_j$ can be fused when all intervening operations commute with $Q_j$,
\[
 \prod_{\mu\in\mathcal F_j}e^{i\vartheta_{\mu,\nu}Q_j}
 =\exp\!\left(iQ_j\sum_{\mu\in\mathcal F_j}\vartheta_{\mu,\nu}\right),
\]
where $\mathcal F_j$ is a group of rotations applied in a single MPS update per shot.
We repeat this fusion after simplifying operators on free $\lvert0\rangle$ qubits, since removing trivial $Z$ factors via $Z\lvert0\rangle=\lvert0\rangle$ can make additional Pauli strings identical.

When a rotation instead acts as $X$ or $Y$ on a free qubit, its multi-qubit coupling can be absorbed into $C$, leaving a local MPS update~\cite{iccr,ofd,clifft}.
For $Q=\sigma_f\otimes\overline P_f$ with $\sigma_f\in\{X,Y\}$ on a free qubit $f$ and $\overline P_f$ on the remaining qubits, the controlled-$\overline P_f$ Clifford gate $D_f$ (with control $f$) obeys
\begin{equation}
 \begin{aligned}
 &e^{i\vartheta Q}(\lvert0\rangle_f\otimes\lvert\eta\rangle)\\
 &\quad=D_f[ e^{i\vartheta\sigma_f}\lvert0\rangle_f
       \otimes\lvert\eta\rangle],\\
 C&\xrightarrow{\rm compile}CD_f,\\
 E(\mathbf b)&\xrightarrow{\rm compile}D_f^\dagger E(\mathbf b)D_f .
 \end{aligned}
 \label{eq:activation-frame}
\end{equation}
Because this identity holds for any state $\lvert\eta\rangle$ of the other qubits, compilation fixes $D_f$ from $Q$ and $f$, while each shot activates $f$ through a single-qubit rotation without increasing the MPS bond dimensions.

Subsequent measurements can release active qubits for reuse~\cite{iccr,clifft}.
For an active Pauli measurement, where the reduced $Q\neq I$ acts only on active qubits (deterministic and free-qubit cases update $C$ and $E$ analytically), the compiler chooses a pivot qubit $r$ in its support and constructs a Clifford $G_r$ satisfying $G_rQG_r^\dagger=Z_r$.
The conditional release map $M_{m,r}=(\lvert0\rangle\langle m\rvert_r\otimes I_{\bar r})G_r$ projects onto the $(-1)^m$ eigenspace of $Q$ and resets $r$ to $\lvert0\rangle$ in the transformed basis,
\begin{equation}
 \begin{aligned}
 C&\xrightarrow{\rm compile}CG_r^\dagger,\\
 E(\mathbf b)&\xrightarrow{\rm compile}G_rE(\mathbf b)G_r^\dagger X_r^m,\\
 \lvert\psi[A_\nu]\rangle&\xrightarrow{\mathrm{shot}\,\nu}
 \frac{M_{m_\nu,r}\lvert\psi[A_\nu]\rangle}{\sqrt{p_{m_\nu}}} .
 \end{aligned}
 \label{eq:measurement-frame}
\end{equation}
Each shot evaluates $\langle Q\rangle$ from its current MPS, samples $m_\nu\in\{0,1\}$ with probability $p_m=[1+(-1)^m\langle Q\rangle]/2$, and substitutes $m=m_\nu$ into $E(\mathbf b)$, with position $r$ now free for later rotations.
Applying the compiled sign correction to the residual eigenvalue $(-1)^{m_\nu}$ recovers the physical measurement outcome.

CAMPS reduces MPS entanglement by changing the Clifford representation~\cite{camps,ofd}.
A Clifford gate $G$ on the stored positions maps the residual state to $\lvert\psi[A'_\nu]\rangle=G\lvert\psi[A_\nu]\rangle$ while leaving the physical state unchanged,
\begin{equation}
 \begin{aligned}
 \lvert\Psi_\nu\rangle
 &=CE_\nu\lvert\psi[A_\nu]\rangle\\
 &=(CG^\dagger)(GE_\nu G^\dagger)\lvert\psi[A'_\nu]\rangle .
 \end{aligned}
 \label{eq:residual-basis-change}
\end{equation}
Rather than applying extra disentangling gates to the MPS at runtime, we realize these basis changes during compilation through the activation and measurement rules, updating later Pauli strings as $Q\xrightarrow{\rm compile}GQG^\dagger$ so that all shots share a common basis.

\begin{figure*}[t]
\begin{minipage}[t]{\resultcolumnwidth}
\vspace{0pt}
\centering
\includegraphics[width=\linewidth]{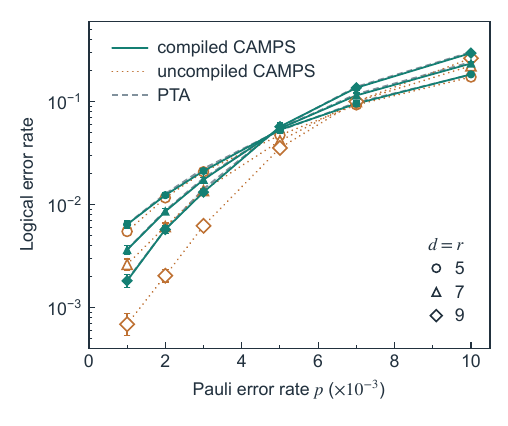}
\caption{Surface-code logical errors under random-sign $ZZ$ noise.
$X$ memory with $r=d$ rounds and independent, equally likely $e^{\pm i\theta ZZ}$ rotations ($\theta=0.02$); compiled and uncompiled CAMPS use $\chi=8$.
Filled and open symbols denote compiled and uncompiled CAMPS; circles, triangles, and diamonds denote $d=5$, $7$, and $9$.
Compiled CAMPS agrees with the exact sign-averaged PTA reference, while uncompiled CAMPS increasingly underestimates logical errors as $d$ grows.
Bars and shading show 95\% confidence intervals.}
\label{fig:surface}
\end{minipage}\hfill
\begin{minipage}[t]{\resultcolumnwidth}
\vspace{0pt}
\centering
\includegraphics[width=\linewidth]{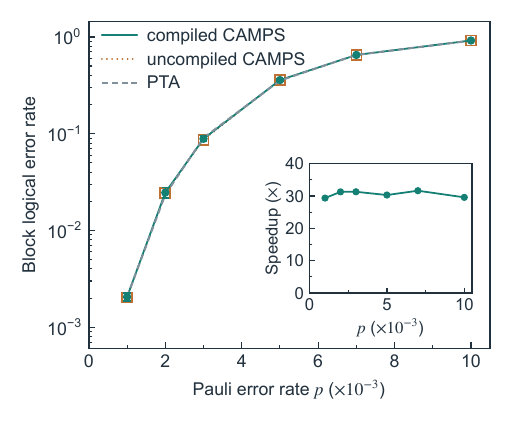}
\caption{BB-code logical errors and compiled CAMPS sampling speedup.
Six-round $X$ memory of the $[[72,12,6]]$ code with independent, equally likely $e^{\pm i\theta Y}$ rotations ($\theta=0.01$); compiled and uncompiled CAMPS use $\chi=4$.
Filled circles and open squares denote compiled and uncompiled CAMPS, both agreeing with the exact sign-averaged PTA reference at this small bond dimension.
Inset: $\bar t_{\mathrm{uncomp}}/\bar t_{\mathrm{comp}}$, showing approximately $30\times$ faster mean sampling with compiled CAMPS at equal $\chi$.
Bars and shading show 95\% confidence intervals.}
\label{fig:bb}
\end{minipage}
\end{figure*}

After fusion, we optimize this shared basis to shorten the MPS intervals spanned by rotations and measurements.
Operations within one block need only a local tensor update, whereas wider operations require contraction and compression across inter-block MPS bonds.
A candidate choice $g$ selects a vacant MPS position for qubit activation in Eq.~\eqref{eq:activation-frame}, or a measurement pivot $r$ where $Q$ acts nontrivially in Eq.~\eqref{eq:measurement-frame}.
Each choice changes the Clifford basis as in Eq.~\eqref{eq:residual-basis-change} and transforms later Pauli strings.
Propagating these strings across a finite lookahead window $\mathcal W$, we minimize the total weighted width,
\begin{equation}
 \begin{aligned}
 \mathcal J(g)&=\sum_{j\in\mathcal W}w_j\Delta_j(g),\\
 g_\star&=\operatorname*{arg\,min}_g\mathcal J(g),
 \end{aligned}
 \label{eq:tensor-layout-cost}
\end{equation}
where $\Delta_j(g)$ is the number of MPS bonds between the first and last blocks affected by operation $j$.
We set $w_j=1$ for rotations and $w_j=2$ for active measurements, since a measurement requires separate contractions to evaluate outcome probabilities and update the conditional state.
Each shot then evolves its MPS using the optimized compiled instructions.
Further details of compiled CAMPS are provided in the Supplemental Material (SM)~\cite{supplemental}.

\emph{Numerical benchmarks.---}We evaluate the accuracy and sampling cost of compiled CAMPS against uncompiled CAMPS~\cite{harper} and physical-qubit MPS baselines on rotated surface codes~\cite{bombin2007,fowler2012} and the $[[72,12,6]]$ bivariate bicycle (BB) code, a high-rate quantum low-density parity-check (qLDPC) code~\cite{bb}.
In uncompiled CAMPS, each shot updates its Clifford operator and $n$-qubit MPS at runtime on the same backend, without qubit recycling, rotation fusion, or basis optimization.
In both codes, coherent Pauli rotations are added to circuit-level Pauli noise models from earlier studies, while all simulators share a decoder built from the Pauli-twirled circuit.
For circuit simulation, however, the Pauli-twirling approximation (PTA) is exact only for symmetric rotation-sign distributions and misses coherent interference in general.

We first simulate rotated surface-code $X$ memory over $r=d$ rounds with coherent $ZZ$ rotations of angle $\theta=0.02$, decoded by minimum-weight perfect matching~\cite{pymatching}.
Following Ref.~\cite{harper}, the Pauli noise model assigns depolarizing probabilities $p_1=0.1p$ and $p_2=p$ after one- and two-qubit Clifford gates, reset error probability $p_R=2p$, and readout error probability $p_M=5p$.
As shown in Fig.~\ref{fig:camps}(c), free-qubit recycling substantially shortens the MPS required for these circuits, reducing $n=49$ physical qubits to at most $K=17$ active qubits at $d=5$ and $n=161$ to $K=47$ at $d=9$.
Evolving the residual state vector gives an exact reference at $K=17$.
At $K=47$, however, a dense state vector would require $2^{47}$ complex amplitudes, whereas compiled CAMPS represents the residual state as an MPS with small bond dimensions.

To test convergence under coherent noise, we set $d=r=5$ and $p=0.005$ with fixed rotation signs and track the logical error rate (LER) $p_L$ versus $\chi$ in Fig.~\ref{fig:camps}(d).
In this fixed-sign setting, coherent interference raises $p_L$ to more than twice the PTA prediction.
Compiled CAMPS agrees with the untruncated reference within sampling uncertainty at $\chi=8$, whereas uncompiled CAMPS requires $\chi=32$ and the physical-qubit MPS remains unconverged through $\chi=64$.
In Fig.~\ref{fig:camps}(e), compiled CAMPS further reduces the median time per shot for state evolution and measurement sampling by more than an order of magnitude relative to uncompiled CAMPS and physical MPS at equal $\chi$.
With the same $K=17$ active space and MPS backend, compiled CAMPS also converges at smaller $\chi$ and samples faster than Clifft-compiled MPS evolution, as shown in Fig.~\ref{fig:compiler-accuracy} of the End Matter.

At larger code distances, $K$ exceeds the reach of state-vector simulation, leaving fixed-sign coherent noise without an exact reference.
To verify accuracy in this regime, we simulate larger surface codes and the BB code under random-sign coherent rotations, drawing each physical sign $s=\pm1$ independently and with equal probability in every shot.
For a Pauli string $P$ and density matrix $\rho$, averaging over $e^{\pm i\theta P}$ gives
\begin{equation}
 \begin{aligned}
 \frac12\sum_{s=\pm1}e^{is\theta P}\rho e^{-is\theta P}
 &=\cos^2\theta\,\rho+\sin^2\theta\,P\rho P.
 \end{aligned}
 \label{eq:sign-averaged-noise}
\end{equation}
Averaging over the independent signs makes PTA exact and allows us to obtain the reference $p_L$ by stabilizer sampling, whereas each compiled or uncompiled CAMPS shot still applies the coherent rotations $e^{is\theta P}$.
Randomizing physical signs adds a sign flip per rotation without changing the compiled Pauli strings $Q_j$ or the MPS basis, because the effective angles in Eq.~\eqref{eq:compiled-rotation} already take fluctuating signs determined by $C$, the stochastic Pauli frame $E_\nu$, and fusion.
At $d=5$, Fig.~\ref{fig:sign-convergence} of the End Matter confirms that compiled CAMPS converges similarly for fixed and random signs, reaching both exact references at $\chi=8$.
Random-sign benchmarks thus provide a reliable accuracy test for general coherent noise.

Figure~\ref{fig:surface} compares the surface-code LER $p_L$ for $d=5$, $7$, and $9$ at $\chi=8$.
Compiled CAMPS agrees closely with the exact PTA reference across all three distances and captures the suppression of logical errors with increasing $d$ at low $p$.
By contrast, uncompiled CAMPS at this bond dimension increasingly underestimates $p_L$ as $d$ grows, with a pronounced deviation at $d=9$ and $p=10^{-3}$.

Finally, we test compiled CAMPS beyond planar connectivity on six-round $X$ memory of the $[[72,12,6]]$ BB code, which reduces qubit overhead at the cost of nonlocal parity checks~\cite{bb}.
Circuit noise includes Pauli errors of uniform probability $p$ on CNOT gates, idling, preparation, and readout, together with independent random-sign $Y$ rotations of angle $\theta=0.01$ on each data qubit per round.
Free-qubit recycling reduces the $n=144$ physical qubits to at most $K=45$ active qubits.
We decode using belief propagation with ordered-statistics postprocessing~\cite{bposd}, and Fig.~\ref{fig:bb} reports the block LER, the probability that at least one of the twelve logical $X$ outcomes is incorrect.
At $\chi=4$, both compiled and uncompiled CAMPS agree well with the exact PTA reference, while the inset of Fig.~\ref{fig:bb} shows that compilation reduces the mean sampling time per shot by a factor of roughly $30$ across the noise scan.

\emph{Conclusion and perspective.---}We have developed compiled CAMPS for repeated QEC sampling under coherent noise beyond Pauli twirling.
Compiling the circuit once lets each shot update a Pauli frame and evolve a residual MPS, while free-qubit recycling, rotation fusion, and basis optimization reduce active qubits, residual entanglement, and operator spans.
On surface- and BB-code benchmarks, compiled CAMPS yields accurate logical error estimates at small $\chi$ and samples over an order of magnitude faster than uncompiled CAMPS at equal $\chi$.

In this work, neither uncompiled nor compiled CAMPS performs iterative disentangler optimization sweeps~\cite{camps} \cite{camps} during MPS evolution. Incorporating such sweeps remains a possible direction for further reducing $\chi$ in deeper circuits.
Mid-circuit feedforward~\cite{shirizly2025} motivates adaptive compilation, while optimized unravelings~\cite{vovk2022,unraveling} may reduce residual MPS entanglement under dissipative noise.
Beyond quantum memories, compiled CAMPS can simulate nonlocal qLDPC logical gates~\cite{williamson2026} as well as magic-state factories~\cite{magna} and cultivation~\cite{cultivation} under coherent hardware noise.
Fast trajectory sampling also makes it practical to calibrate decoders beyond Pauli priors and train machine-learning decoders~\cite{bausch2024} on coherent errors.

\textbf{Acknowledgments:}
We thank Riling Li and Yiming Zhang for helpful discussions. The computations in this work were run on the Siyuan-1 cluster supported by the Center for High Performance Computing at Shanghai Jiao Tong University. M.Q. acknowledges support from the National Natural Science Foundation of China (Grants No. 12522406 and No. 12274290), and the National Key Research and Development Program of MOST of China (2022YFA1405400). We acknowledge the use of ChatGPT for assistance with code development and manuscript editing. The authors take full responsibility for the content and conclusions of this work.
\bibliography{references}
\clearpage
\onecolumngrid
\begin{center}
\textbf{End Matter}
\end{center}
\twocolumngrid
\raggedbottom
\input{end_matter}
\end{document}


\title{Supplemental Material: Accelerating Quantum Error Correction Simulations under Coherent Noise with Compiled Clifford-Augmented Matrix Product States}
\author{Jiale Huang}
\affiliation{Key Laboratory of Artificial Structures and Quantum Control (Ministry of Education), School of Physics and Astronomy, Shanghai Jiao Tong University, Shanghai 200240, China}
\affiliation{FieldQuantum Research Institute, Beijing 100176, China}
\affiliation{FieldQuantum Frontier Research Institute, Beijing 100190, China}

\author{Yaojian Chen}
\affiliation{FieldQuantum Research Institute, Beijing 100176, China}
\affiliation{FieldQuantum Frontier Research Institute, Beijing 100190, China}

\author{Ke Liu}
\affiliation{Hefei National Research Center for Physical Sciences at the Microscale and School of Physical Sciences, University of Science and Technology of China, Hefei 230026, China}
\affiliation{Shanghai Research Center for Quantum Science and CAS Center for Excellence in Quantum Information and Quantum Physics, University of Science and Technology of China, Shanghai 201315, China}

\author{Dingshun Lv}
\email[Corresponding author: ]{lvdingshun@fieldquanta.com}
\affiliation{FieldQuantum Research Institute, Beijing 100176, China}
\affiliation{FieldQuantum Frontier Research Institute, Beijing 100190, China}

\author{Mingpu Qin}
\email[Corresponding author: ]{qinmingpu@sjtu.edu.cn}
\affiliation{Key Laboratory of Artificial Structures and Quantum Control (Ministry of Education), School of Physics and Astronomy, Shanghai Jiao Tong University, Shanghai 200240, China}
\maketitle

\begingroup
\small
\pdfbookmark[1]{Contents}{smcontents}
{\centering\bfseries CONTENTS\par}
\nobreak\vspace{0.45\baselineskip}
\hypersetup{linktoc=all}
\makeatletter
\renewcommand*{\l@section}[2]{%
  \addvspace{2.4pt}%
  \@dottedtocline{1}{0pt}{2.5em}{\bfseries #1}{\bfseries #2}}
\renewcommand*{\l@subsection}[2]{%
  \@dottedtocline{2}{2.5em}{1.8em}{#1}{#2}}
\@starttoc{toc}
\makeatother
\endgroup
\par\addvspace{\baselineskip}

We compile a noisy circuit once for simulation with a compiled Clifford-augmented matrix product state (CAMPS). The compiler fixes the Pauli axes of rotations and measurements and the rules for their signs. Each shot samples the noise, evolves the stored matrix product state (MPS), and produces a measurement record for decoding.
The circuits have fixed gate locations and can include Clifford gates, Pauli rotations, stochastic Pauli faults, measurements, resets, and parity-controlled Pauli feedback.
Our rotation convention is
\[
R_P(\theta)\equiv e^{i\theta P}=\cos\theta\,I+i\sin\theta\,P,
\qquad R_X(\theta)=e^{i\theta X},\qquad R_Z(\theta)=e^{i\theta Z}.
\]
Here $P$ is a Hermitian Pauli string and $\theta\in\mathbb R$ is the rotation angle. Operator products act from right to left.
Sections~\ref{sec:representation} and~\ref{sec:active} derive the update rules. Section~\ref{sec:fusion} builds the shared instruction list, Sec.~\ref{sec:mps} executes it for one shot, and Sec.~\ref{sec:protocol} describes the noise models and how logical failures are counted.
Two qubits per MPS block are used throughout, except in the block-size comparison.

\newpage
\section*{Notation}
\begingroup
\small
\renewcommand{\arraystretch}{1.08}
\noindent
\begin{tabular}{@{}ll@{}}
\hline\hline
\smcell{0.30}{Symbol} & \smcell{0.67}{Definition}\\
\hline
\smcell{0.30}{$n$} & \smcell{0.67}{Total number of physical circuit qubits, including ancillas.}\\
\smcell{0.30}{$k(t),K=\max_t k(t)$} & \smcell{0.67}{Active-qubit count and its peak; the $K$ stored positions hold active qubits and free qubits in $\lvert0\rangle$.}\\
\smcell{0.30}{$C(t)$} & \smcell{0.67}{Compiled $n$-qubit Clifford, shared by all shots.}\\
\smcell{0.30}{$E(t;\mathbf b),E_\nu(t)$} & \smcell{0.67}{Symbolic Pauli frame and its value $E(t;\mathbf b_\nu)$ in shot $\nu$.}\\
\smcell{0.30}{$\mathbf b_\nu$} & \smcell{0.67}{Bits for Pauli faults, readout errors, physical rotation signs when sampled, and residual measurement outcomes.}\\
\smcell{0.30}{$\lvert\Psi_\nu\rangle$} & \smcell{0.67}{Full physical state, $C E_\nu\lvert\psi[A_\nu]\rangle$.}\\
\smcell{0.30}{$\lvert\psi[A_\nu]\rangle,\lvert\phi_\nu\rangle$} & \smcell{0.67}{Full residual state and stored $K$-qubit state, related by $\lvert\psi[A_\nu]\rangle=\lvert\phi_\nu\rangle\otimes\lvert0\rangle^{\otimes(n-K)}$.}\\
\smcell{0.30}{$A_\nu$} & \smcell{0.67}{MPS tensors evolved numerically in shot $\nu$.}\\
\smcell{0.30}{$\chi$} & \smcell{0.67}{Maximum retained MPS bond dimension between blocks.}\\
\smcell{0.30}{$P_j,Q_j$} & \smcell{0.67}{Physical Pauli string and compiled unsigned residual Pauli string.}\\
\smcell{0.30}{$\theta_j,\vartheta_{j,\nu}$} & \smcell{0.67}{Physical rotation angle and effective residual angle in shot $\nu$, including compiled signs and fusion.}\\
\smcell{0.30}{$\Delta_j$} & \smcell{0.67}{Number of MPS bonds between the first and last blocks on which residual operator $j$ acts nontrivially.}\\
\hline\hline
\end{tabular}
\par
\endgroup

\section{Compiled CAMPS representation}
\label{sec:representation}
\subsection{State decomposition}
At circuit step $t$, compiled CAMPS represents the physical state in shot $\nu$ as~\cite{camps,clifft,symft}
\begin{equation}
 \begin{aligned}
 \lvert\Psi_\nu(t)\rangle&=C(t)E_\nu(t)\lvert\psi[A_\nu(t)]\rangle,\\
 E_\nu(t)&=E(t;\mathbf b_\nu),\\
 \lvert\psi[A_\nu(t)]\rangle&=\lvert\phi_\nu(t)\rangle\otimes\lvert0\rangle^{\otimes(n-K)}.
 \end{aligned}
 \label{eq:compiled-state}
\end{equation}
The Clifford $C(t)$ is shared by all shots. The Pauli frame $E(t;\mathbf b)$ stores stochastic faults and measurement corrections as functions of event bits $\mathbf b$.
Each shot supplies its own record $\mathbf b_\nu$ and evolves its own MPS tensors $A_\nu$.

A \emph{free qubit} is a residual qubit in the known state $\lvert0\rangle$, unentangled with the other residual qubits.
The compiler marks qubits as active when they enter the numerical evolution and releases them after suitable measurements.
It reserves $K=\max_t k(t)$ MPS positions for the peak active count: at step $t$, these hold $k(t)$ active qubits and $K-k(t)$ free qubits.
The remaining $n-K$ free qubits are represented by the final product state in Eq.~\eqref{eq:compiled-state}.
These labels refer to the residual state; the physical state is reconstructed by $CE_\nu$.

\subsection{Pauli-frame updates and rotation signs}
Start from $C=E=I$.
A Clifford gate $U$ gives $UCE=(UC)E$, while a Pauli fault obeys $F^eC=C(C^\dagger FC)^e$.
The updates are therefore
\[
\begin{array}{c|c}
\text{Physical operation}&\text{Compiled update}\\ \hline
\text{Clifford }U&C\mapsto UC\\
\text{Pauli fault }F^e&E\mapsto(C^\dagger FC)^eE.
\end{array}
\]
The bit $e=0$ denotes no fault and $e=1$ denotes a fault; it is sampled for each trajectory.
Successive Pauli components combine by exclusive OR (XOR), denoted by $\oplus$. For example,
\[
Z_1^{b_2}Z_1^{b_3}=Z_1^{b_2\oplus b_3}.
\]
If both events occur, the two $Z_1$ operators cancel; if only one occurs, a $Z_1$ remains.
Clifford conjugation maps Pauli strings to Pauli strings. Thus, up to global phase, the frame can be stored as
\[
E(\mathbf b)=\prod_rX_r^{x_r(\mathbf b)}Z_r^{z_r(\mathbf b)},
\]
where each binary exponent is an XOR of selected event bits and a possible fixed bit.
Parity-controlled Pauli feedback has the same form.

\emph{Pauli-frame signs in residual rotations.---}Clifford conjugation fixes the residual axis: $C^\dagger P_jC=\pm Q_j$.
The Pauli frame $E$ preserves this rotation when $EQ_j=Q_jE$ and reverses its angle when $EQ_j=-Q_jE$.
For a single-qubit $X$ rotation,
\[
X^\dagger R_X(\theta)X=R_X(\theta),\qquad
Z^\dagger R_X(\theta)Z=R_X(-\theta),
\]
because $X^\dagger XX=X$ and $Z^\dagger XZ=-X$.
For example, take $E=X_1^{b_1}Z_1^{b_2\oplus b_3}$.
Its $X_1$ component commutes with the rotation, while its $Z_1$ component gives
\[
E^\dagger R_{X_1}(\theta)E=
\begin{cases}
R_{X_1}(+\theta),&b_2=b_3,\\
R_{X_1}(-\theta),&b_2\ne b_3.
\end{cases}
\]
This example extends to any Pauli string:
\begin{align}
E_\nu^\dagger C^\dagger R_{P_j}(\theta_j)CE_\nu
 &=R_{Q_j}(\vartheta_{j,\nu}),\qquad
 \vartheta_{j,\nu}=(-1)^{\ell_j(\mathbf b_\nu)}\theta_j,
 \label{eq:compiled-rotation}\\
\ell_j(\mathbf b_\nu)&=c_j\oplus\bigoplus_\mu h_{j\mu}b_{\nu,\mu}.
 \label{eq:compiled-sign}
\end{align}
Here $h_{j\mu}=1$ selects an event bit that reverses the rotation's sign, and $c_j$ includes fixed signs from the Clifford and frame.
In the example above, $c_j=0$ and $(h_{j1},h_{j2},h_{j3})=(0,1,1)$: $b_1$ has no effect on the angle.
Compilation fixes these coefficients. Each shot substitutes its event bits, obtaining $+\theta_j$ for $\ell_j=0$ and $-\theta_j$ for $\ell_j=1$.
For a physical rotation $R_{P_j}(s_j\theta_j)$, the noise model supplies $s_j$ and the Clifford and Pauli frame determine $\ell_j$. The residual angle in shot $\nu$ is $s_j(-1)^{\ell_j(\mathbf b_\nu)}\theta_j$.

\begin{figure}[ht]
\centering
\includegraphics{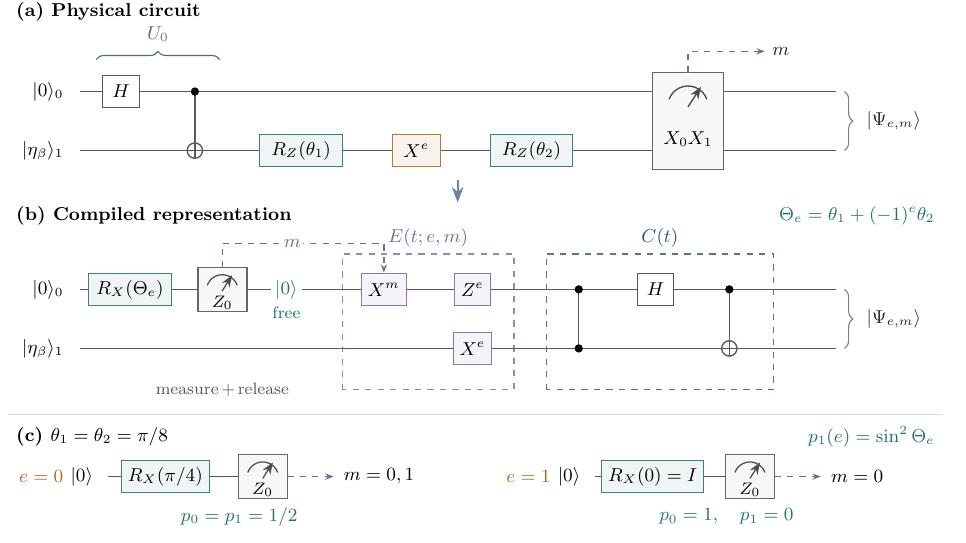}
\caption{The two-qubit example, with gates read from left to right. (a) Physical rotations and a Pauli fault, followed by a joint $X_0X_1$ measurement. (b) The corresponding residual rotation, $Z_0$ measurement, and release. The dashed arrow writes $m$ into the Pauli frame; the dashed boxes reconstruct the conditional physical state using $E(t;e,m)$ and the shared Clifford $C(t)$. (c) Measurement probabilities for the two fault branches at $\theta_1=\theta_2=\pi/8$.}
\label{fig:sm-workflow}
\end{figure}

\begingroup
\setlength{\abovedisplayskip}{5pt}
\setlength{\belowdisplayskip}{5pt}
\setlength{\abovedisplayshortskip}{0pt}
\setlength{\belowdisplayshortskip}{5pt}
\subsection{Two-qubit example with measurement}
Figure~\ref{fig:sm-workflow} starts from
\[
\lvert0\rangle_0\otimes\lvert\eta_\beta\rangle_1,
\qquad \lvert\eta_\beta\rangle=R_X(\beta)\lvert0\rangle
=\cos\beta\lvert0\rangle+i\sin\beta\lvert1\rangle,
\qquad U_0=\operatorname{CNOT}_{0\to1}H_0.
\]
The state just before the physical $X_0X_1$ measurement in panel (a) is
\[
\lvert\Psi_e\rangle=R_{Z_1}(\theta_2)X_1^eR_{Z_1}(\theta_1)U_0
 (\lvert0\rangle_0\otimes\lvert\eta_\beta\rangle_1).
\]
Conjugation by $U_0$ gives the residual axes and the fused angle:
\[
\begin{aligned}
U_0^\dagger Z_1U_0&=X_0Z_1\equiv Q,&U_0^\dagger X_1U_0&=X_1,\\
R_Q(\theta_2)X_1^eR_Q(\theta_1)&=X_1^eR_Q(\Theta_e),&
\Theta_e&=\theta_1+(-1)^e\theta_2.
\end{aligned}
\]
Activating qubit $0$ absorbs $\operatorname{CZ}_{01}$ into the Clifford and leaves a single-qubit residual rotation:
\[
\begin{aligned}
C_*&=U_0\operatorname{CZ}_{01},\qquad E_*(e)=(Z_0X_1)^e,\\
\lvert\Psi_e\rangle&=C_*E_*(e)\lvert\psi_e\rangle,\qquad
\lvert\psi_e\rangle=R_{X_0}(\Theta_e)\lvert0\rangle_0\otimes\lvert\eta_\beta\rangle_1.
\end{aligned}
\]
Now measure $X_0X_1$ with eigenvalue $(-1)^m$, where $m=0,1$.
The Clifford maps this operator to $Z_0$, and the frame preserves its sign:
\[
C_*^\dagger(X_0X_1)C_*=Z_0,\qquad E_*^\dagger Z_0E_*=Z_0,
\qquad p_0(e)=\cos^2\Theta_e,\quad p_1(e)=\sin^2\Theta_e.
\]
The residual measurement leaves qubit $0$ in $\lvert m\rangle$.
Returning it to the free state $\lvert0\rangle$ transfers its value into the frame as $X_0^m$:
\[
\begin{aligned}
E_{\rm after}(e,m)&=E_*(e)X_0^m,\qquad
\lvert\psi_{e,m}\rangle=\lvert0\rangle_0\otimes\lvert\eta_\beta\rangle_1,\\
\lvert\Psi_{e,m}\rangle&\simeq C_*E_{\rm after}(e,m)\lvert\psi_{e,m}\rangle.
\end{aligned}
\]
Here $\simeq$ omits a branch-global phase. These are the dashed boxes in panel (b):
$C(t)=C_*$ and $E(t;e,m)=E_{\rm after}(e,m)$.
For $\theta_1=\theta_2=\pi/8$, $e=0$ gives $\Theta_0=\pi/4$ and $p_0=p_1=1/2$, whereas $e=1$ gives $\Theta_1=0$ and $m=0$ with probability one [panel (c)].
Section~\ref{sec:active} derives the activation and release identities and continues the example with qubit reuse.
\par\endgroup

\section{Qubit activation, measurement, and reuse}
\label{sec:active}
A free qubit is in $\lvert0\rangle$ and unentangled with the other residual qubits.
A rotation can activate it; a later measurement can make an active qubit free again.
The identities below describe these two changes of representation and determine the active count $k(t)$.
Figure~\ref{fig:sm-reuse} illustrates them with two-qubit circuits.

\subsection{Activating a free qubit}
Consider a rotation whose Pauli string contains $X$ or $Y$ on a free qubit $f$:
$Q=\sigma_f\otimes\overline P_f$, with $\sigma_f=X$ or $Y$.
The other qubits may be in any state $\lvert\eta\rangle$.
Define $D_f$ as a controlled-$\overline P_f$ gate with control qubit $f$:
\begin{equation}
D_f=\lvert0\rangle\langle0\rvert_f\otimes I
 +\lvert1\rangle\langle1\rvert_f\otimes\overline P_f.
\label{eq:promotion-clifford}
\end{equation}
\noindent\makebox[\linewidth][c]{\includegraphics{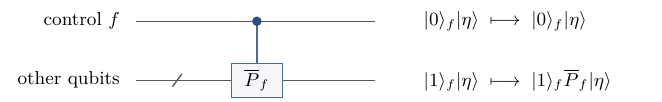}}\par
The rotation creates two terms,
\[
R_Q(\vartheta)\lvert0\rangle_f\lvert\eta\rangle
=\cos\vartheta\lvert0\rangle_f\lvert\eta\rangle
 +i\sin\vartheta\,\sigma_f\lvert0\rangle_f\overline P_f\lvert\eta\rangle.
\]
The two branches in the diagram reproduce these terms: $D_f$ applies $\overline P_f$ only when $f$ is in $\lvert1\rangle$.
Thus a local rotation followed by this Clifford gives the same state~\cite{ofd,iccr}:
\begin{equation}
R_Q(\vartheta)(\lvert0\rangle_f\otimes\lvert\eta\rangle)
=D_f\bigl[R_{\sigma_f}(\vartheta)\lvert0\rangle_f\otimes\lvert\eta\rangle\bigr].
\label{eq:promotion-identity}
\end{equation}
For example, $Q=X_0Z_1$ gives $D_0=\operatorname{CZ}_{01}$ [Fig.~\ref{fig:sm-reuse}(a)]:
\[
\cos\vartheta\lvert0\rangle_0\lvert\eta\rangle_1
 +i\sin\vartheta\lvert1\rangle_0Z_1\lvert\eta\rangle_1
=\operatorname{CZ}_{01}
 \bigl[(\cos\vartheta\lvert0\rangle_0+i\sin\vartheta\lvert1\rangle_0)
 \otimes\lvert\eta\rangle_1\bigr].
\]
Absorbing $D_f$ into the compiled representation gives
\[
\begin{aligned}
C'&=CD_f,& E'&=D_f^\dagger ED_f,\\
\lvert\psi'\rangle&=R_{\sigma_f}(\vartheta)\lvert0\rangle_f\otimes\lvert\eta\rangle,
& k'&=k+1.
\end{aligned}
\]
The stored state remains a product across $f$ and the other qubits, so activating $f$ changes only its local tensor and leaves the virtual bond dimensions unchanged.
For $X$ the new vector is $\cos\vartheta\lvert0\rangle+i\sin\vartheta\lvert1\rangle$; for $Y$ it is $\cos\vartheta\lvert0\rangle-\sin\vartheta\lvert1\rangle$.
The compiler assigns $f$ to a free MPS position and includes this relabeling in $C$.

If a rotation has only $I$ or $Z$ on free qubits, their known action is removed using $Z\lvert0\rangle=\lvert0\rangle$.
The remaining rotation acts on the active qubits through the matrix product operator (MPO) update in Sec.~\ref{sec:mps}; if its Pauli string is the identity, it contributes a global phase.

\subsection{Measurement cases and qubit release}
For an unsigned residual Pauli $Q$, outcome $m$ labels the eigenvalue $(-1)^m$:
\begin{equation}
\Pi_m(Q)=\frac{I+(-1)^mQ}{2},\qquad
p_m=\frac{1+(-1)^m\langle Q\rangle}{2},\qquad m\in\{0,1\}.
\label{eq:born-projector}
\end{equation}
First remove any $Z$ operators on free qubits using $Z\lvert0\rangle=\lvert0\rangle$.
The remaining Pauli operators determine which of the following rules applies.
\begin{center}
\begin{tabular}{@{}cllr@{}}
\hline\hline
Case & \smcell{0.34}{Where $Q$ acts} & \smcell{0.39}{Outcome and state update} & Active count\\ \hline
1 & \smcell{0.34}{Identity} & \smcell{0.39}{$m=0$; state unchanged} & $k\to k$\\
2 & \smcell{0.34}{$X$ or $Y$ on a free qubit} & \smcell{0.39}{$p_0=p_1=1/2$; update $C$ and $E$} & $k\to k$\\
3 & \smcell{0.34}{$X$, $Y$, or $Z$ only on active qubits} & \smcell{0.39}{Evaluate $\langle Q\rangle$; sample $m$ and release one qubit} & $k\to k-1$\\
\hline\hline
\end{tabular}
\end{center}
These are outcomes of the unsigned residual operator; conversion to the physical outcome via the compiled sign is described at the end of this section.

\emph{Case 1: a deterministic outcome.---}The input is already a $+1$ eigenstate of $Q$.
For example, $Q=Z_f$ with $f$ free gives $Q\lvert\psi\rangle=\lvert\psi\rangle$.
Thus $p_0=1$, the projection leaves the state unchanged, and $k'=k$.

\emph{Case 2: $X$ or $Y$ on a free qubit.---}Write $Q=\sigma_f\otimes\overline P_f$ with $f$ free.
The known state of $f$ fixes the expectation value:
\[
\langle Q\rangle
=\langle0\rvert\sigma_f\lvert0\rangle\,
 \langle\eta\rvert\overline P_f\lvert\eta\rangle=0.
\]
Equation~\eqref{eq:born-projector} therefore gives $p_0=p_1=1/2$ for any $\lvert\eta\rangle$.
For $Q=X_f$, the normalized state of the measured qubit is $HX^m\lvert0\rangle$. The same $H$ applies for both outcomes.
For a general $\overline P_f$, the same construction uses the controlled Pauli $D_f$ from Eq.~\eqref{eq:promotion-clifford}:
\begin{equation}
\Pi_m(Q)(\lvert0\rangle_f\otimes\lvert\eta\rangle)
=\frac{1}{\sqrt2}D_fH_{\sigma,f}X_f^m
 (\lvert0\rangle_f\otimes\lvert\eta\rangle),
\label{eq:fair-measurement}
\end{equation}
where $H_X=H$, $H_Y=SH$, and $S=\operatorname{diag}(1,i)$.
Writing $F_f=D_fH_{\sigma,f}$, the updates are
\[
C'=CF_f,\qquad E'=F_f^\dagger EF_fX_f^m,\qquad
\lvert\psi'\rangle=\lvert\psi\rangle.
\]
The compiler fixes $F_f$ for both outcomes. The sampled bit $m$ enters the frame through $X_f^m$. The residual qubit $f$ stays free, so $k'=k$~\cite{iccr,clifft}.

\begin{figure}[!ht]
\centering
\includegraphics{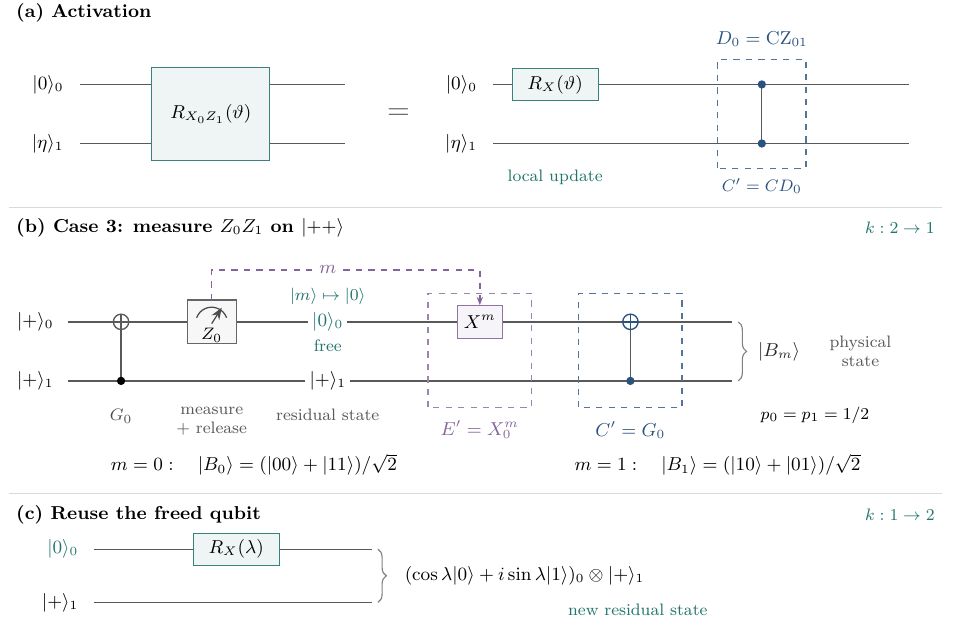}
\caption{Qubit activation and reuse in explicit circuits. (a) The rotation on $\lvert0\rangle_0\lvert\eta\rangle_1$ becomes a local $R_X(\vartheta)$ and a controlled $Z$ absorbed into $C$. (b) Case 3 for $Q=Z_0Z_1$, with input $\lvert++\rangle$ and $C=E=I$. The first CNOT moves the parity to qubit $0$; measurement and release leave the residual state $\lvert0+\rangle$. The dashed boxes, $E'=X_0^m$ and $C'=G_0$, reconstruct the two physical parity states shown below the circuit. (c) The freed qubit is used by a later residual $R_X(\lambda)$. The displayed product state is the new residual state.}
\label{fig:sm-reuse}
\end{figure}

\emph{Case 3: measurement on active qubits.---}Here the remaining $Q$ is nonidentity and acts only on active qubits.
The compiler chooses a pivot $r$ where $Q$ acts as $X$, $Y$, or $Z$ (Sec.~\ref{sec:fusion}).
Apply $H$ at each $X$ and $HS^\dagger$ at each $Y$, so that only $I$ and $Z$ remain.
For each other qubit carrying $Z$, apply a CNOT from that qubit to $r$.
Together these gates form the Clifford $G_r$ with $G_rQG_r^\dagger=Z_r$.

Each shot computes $\langle Q\rangle$ in its stored state and draws $m$ using Eq.~\eqref{eq:born-projector}. In the basis fixed by $G_r$, projection leaves $r$ in $\lvert m\rangle$.
Returning it to $\lvert0\rangle$ defines the release map
\begin{equation}
\begin{aligned}
M_{m,r}&=(\lvert0\rangle\langle m\rvert_r\otimes I_{\bar r})G_r,\\
\lvert\phi'\rangle&=\frac{M_{m,r}\lvert\phi\rangle}{\sqrt{p_m}},
\qquad M_{m,r}^\dagger M_{m,r}=\Pi_m(Q),\qquad k'=k-1.
\end{aligned}
\label{eq:measurement-release}
\end{equation}
The output has the form $\lvert0\rangle_r\otimes\lvert\eta_m\rangle$.
Store the outcome in $E'$ and absorb the inverse basis change into $C'$:
\[
C'=CG_r^\dagger,\qquad E'=G_rEG_r^\dagger X_r^m.
\]
These updates reconstruct the projected physical state, since
\[
C'E'M_{m,r}=CE\Pi_m(Q).
\]
The remaining state $\lvert\eta_m\rangle$ can depend on the outcome.
An active qubit has become free, giving the $k'=k-1$ update in the third row of the table.

\emph{A parity-measurement example.---}Take $Q=Z_0Z_1$, choose $r=0$, and write the normalized residual input as
$\lvert\phi\rangle=\sum_{u,v=0}^1a_{uv}\lvert u\rangle_0\lvert v\rangle_1$.
Here $G_0=\operatorname{CNOT}_{1\to0}$ computes the parity on qubit $0$:
\[
\lvert u,v\rangle\xrightarrow{\ G_0\ }\lvert u\oplus v,v\rangle,
\qquad G_0(Z_0Z_1)G_0^\dagger=Z_0.
\]
Measuring $Z_0$ selects $u\oplus v=m$, so
\[
\begin{aligned}
M_{m,0}\lvert\phi\rangle
 &=\lvert0\rangle_0\otimes
 \bigl(a_{m,0}\lvert0\rangle_1+a_{m\oplus1,1}\lvert1\rangle_1\bigr),\\
p_m&=|a_{m,0}|^2+|a_{m\oplus1,1}|^2,\qquad
\lvert\eta_m\rangle_1
=\frac{a_{m,0}\lvert0\rangle_1+a_{m\oplus1,1}\lvert1\rangle_1}{\sqrt{p_m}}.
\end{aligned}
\]
For example, $\lvert\phi\rangle=\lvert+\rangle_0\lvert+\rangle_1$, with $\lvert+\rangle=(\lvert0\rangle+\lvert1\rangle)/\sqrt2$, gives $p_0=p_1=1/2$ and $\lvert\eta_m\rangle=\lvert+\rangle$.
The released residual state is $\lvert0+\rangle$.
For initial $C=E=I$, the updated Clifford and frame reconstruct the two parity eigenstates:
\[
\operatorname{CNOT}_{1\to0}X_0^m\lvert0+\rangle
=\begin{cases}
(\lvert00\rangle+\lvert11\rangle)/\sqrt2,&m=0,\\
(\lvert10\rangle+\lvert01\rangle)/\sqrt2,&m=1.
\end{cases}
\]
Figure~\ref{fig:sm-reuse}(b) shows this $\lvert++\rangle$ example: the residual state becomes $\lvert0+\rangle$, while $C'=G_0$ and $E'=X_0^m$ reconstruct the even- or odd-parity state.

\subsection{Qubit reuse and physical readout}
After release, a later residual rotation containing $X_r$ or $Y_r$ can activate the same position by Eq.~\eqref{eq:promotion-identity} [Fig.~\ref{fig:sm-reuse}(c)].
In the example of Fig.~\ref{fig:sm-workflow}, $Q=Z_0$ already, so $G_0=I$ and
\[
M_{m,0}\lvert\psi_e\rangle
=a_m(e)\lvert0\rangle_0\lvert\eta_\beta\rangle_1,
\qquad a_0(e)=\cos\Theta_e,\qquad a_1(e)=i\sin\Theta_e.
\]
The frame becomes $E_{\rm after}=E_*X_0^m$.
A subsequent physical $R_{Z_1}(\lambda)$ then acts on the freed qubit because
\[
C_*^\dagger Z_1C_*=X_0,\qquad
E_{\rm after}^\dagger X_0E_{\rm after}=(-1)^eX_0,
\qquad\lvert0\rangle_0\mapsto R_{X_0}((-1)^e\lambda)\lvert0\rangle_0.
\]
Including the preparation of $\lvert\eta_\beta\rangle$, the active count is
\[
0\xrightarrow{R_{X_1}(\beta)}1
\xrightarrow{\text{activation}}2
\xrightarrow{\text{release}}1
\xrightarrow{\text{reuse}}2,\qquad K=2.
\]
This schedule is compiled once and shared by all shots, including branches with $\Theta_e=0$.

\emph{Physical outcomes and reset.---}For a physical measurement of $P$, the compiled sign obeys
$E^\dagger C^\dagger PCE=(-1)^{\ell_j}Q$.
The parity $\ell_j$ converts the residual outcome $m$ to the physical outcome $m_{\rm phys}$. The readout error bit $e_M$ determines the recorded result $y$,
\[
m_{\rm phys}=m\oplus\ell_j,\qquad y=m_{\rm phys}\oplus e_M.
\]
Classical feedback uses $y$.
A physical reset after a $Z$ measurement applies $X^{m_{\rm phys}}$, followed by any modeled preparation error.
The residual release in Eq.~\eqref{eq:measurement-release} frees an MPS position while $C'$ and $E'$ retain the measured physical state; a circuit reset adds the physical correction just described.

\section{Rotation fusion and operator-width optimization}
\label{sec:fusion}
After transporting the Clifford and Pauli frame as in Sec.~\ref{sec:representation}, the compiler fuses rotations about the same Pauli axis. Applying the free-qubit rules of Sec.~\ref{sec:active} can make two previously different axes equal, so the compiler fuses again. It then chooses activation positions and measurement pivots to shorten the MPS intervals of later operations.

\begingroup
\setlength{\abovedisplayskip}{6pt}
\setlength{\belowdisplayskip}{6pt}
\setlength{\abovedisplayshortskip}{0pt}
\setlength{\belowdisplayshortskip}{6pt}
\subsection{First fusion and reordering}
For a group $\mathcal F_j$ of rotations about the same unsigned axis $Q_j$, the signed angles add. Write the physical angle of rotation $\mu$ as $s_\mu\theta_\mu$, with $s_\mu=1$ when the circuit specifies $\theta_\mu$ directly. Then
\begin{equation}
\prod_{\mu\in\mathcal F_j}R_{Q_j}\!\left(s_\mu\theta_\mu(-1)^{\ell_\mu(\mathbf b_\nu)}\right)
 =R_{Q_j}(\vartheta_{j,\nu}),\qquad
\vartheta_{j,\nu}=\sum_{\mu\in\mathcal F_j}
 s_\mu\theta_\mu(-1)^{\ell_\mu(\mathbf b_\nu)}.
\label{eq:fused-angle}
\end{equation}
The compiler adds the sign bit $(1-s_\mu)/2$ to $\ell_\mu$ modulo two and collects terms with the same parity, as in
$\alpha(-1)^e+\gamma(-1)^e=(\alpha+\gamma)(-1)^e$.
Each shot evaluates the resulting sum once.

\emph{Reordering before fusion.---}Rotations with the same axis may be separated by other operations.
The compiler first brings them together so that Eq.~\eqref{eq:fused-angle} can be applied.
For example, if $[P,T]=0$,
\[
R_P(\alpha)R_T(\beta)R_P(\gamma)
 =R_T(\beta)R_P(\alpha)R_P(\gamma)
 =R_T(\beta)R_P(\alpha+\gamma).
\]
Moving one $P$ rotation past the intervening $T$ rotation therefore reduces three rotations to two.
Each such exchange must satisfy two conditions.

\emph{(i) The quantum operations commute.}
For example, $P=X_0X_1$ and $T=Z_0Z_1$ commute: exchanging $X$ and $Z$ contributes one minus sign on each qubit, and the two signs cancel.
For the general check, encode the local Paulis as
$I:(0,0)$, $X:(1,0)$, $Z:(0,1)$, and $Y:(1,1)$, and collect these bits into
$P=(\mathbf x,\mathbf z)$ and $T=(\mathbf x',\mathbf z')$.
The commutation test is
\[
\omega(P,T)=\mathbf x\cdot\mathbf z'
             +\mathbf z\cdot\mathbf x'=0\pmod2.
\]
This sum counts local anticommutations modulo two; for $X_0X_1$ and $Z_0Z_1$, it gives $1+1=0\pmod2$.
For a measurement of $T$, the same condition gives
$R_P(\alpha)\Pi_m(T)=\Pi_m(T)R_P(\alpha)$ for each outcome $m$.

\emph{(ii) Classical bit dependencies are respected.}
Consider the execution order
\[
\underbrace{\text{measure }Z_2\ \longrightarrow\ m}_{\text{produces }m}
\quad\longrightarrow\quad
\underbrace{R_{X_0}\!\left((-1)^m\alpha\right)}_{\text{uses }m}.
\]
Here $X_0$ and $Z_2$ commute, but the rotation must follow the measurement:
$m$ selects the angle $+\alpha$ or $-\alpha$.
To check this dependence, let $\mathcal R_i$ be the classical bits read by instruction $i$, and $\mathcal W_i$ the bits it produces or overwrites.
Exchanging instructions $i$ and $j$ requires
\[
\mathcal W_i\cap(\mathcal R_j\cup\mathcal W_j)=\varnothing,
\qquad \mathcal W_j\cap\mathcal R_i=\varnothing.
\]
Neither instruction may read or overwrite a bit written by the other.
In the example, $\mathcal W_{\rm meas}\cap\mathcal R_{\rm rot}=\{m\}$, so the exchange fails this test.
The same two checks allow measurement blocks to move earlier across eligible rotations, which can enable earlier qubit release.
The measurements retain their order relative to one another.

\subsection{Second fusion after free-qubit reduction}
Let $f$ be free and qubits $0,1$ be active, with input $\lvert0\rangle_f\otimes\lvert\eta\rangle_{01}$.
Set $P=Z_fX_0X_1$, $T=Z_0Z_1$, and $Q=X_0X_1$, and consider
\[
U_e=R_P(\alpha)R_T(\beta)R_P((-1)^e\gamma)R_Q(\delta),
\]
where $e$ is an event bit in the Pauli frame.
Figure~\ref{fig:sm-two-fusions} follows this circuit through both fusion passes.

\begin{figure}[!h]
\centering
\includegraphics[width=\linewidth]{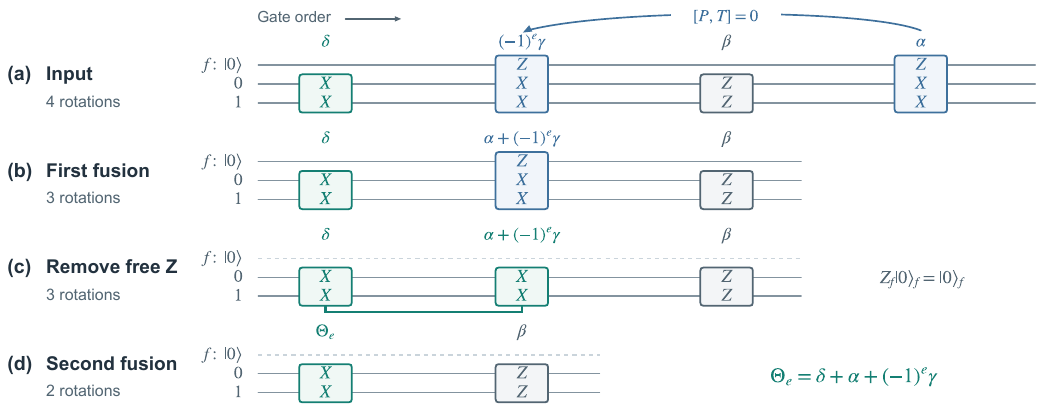}
\caption{Two equal-axis fusion passes. Each gate encloses the Pauli factors of one rotation, with its angle above. Gates act from left to right. (a,b) The two blue rotations commute with the intervening $Z_0Z_1$ rotation and fuse. (c) Since $Z_f\lvert0\rangle_f=\lvert0\rangle_f$, both bracketed rotations now have axis $X_0X_1$. (d) Their angles add to $\Theta_e$, leaving two rotations.}
\label{fig:sm-two-fusions}
\end{figure}

Since $[P,T]=0$, the first pass combines the two $P$ rotations into $R_P(\alpha+(-1)^e\gamma)$. On the free qubit, $Z_f\lvert0\rangle_f=\lvert0\rangle_f$, so this rotation now has the same axis $X_0X_1$ as $R_Q(\delta)$. The second pass adds $\delta$, giving $\Theta_e=\alpha+(-1)^e\gamma+\delta$ and
\[
U_e(\lvert0\rangle_f\otimes\lvert\eta\rangle_{01})
 =\lvert0\rangle_f\otimes
 R_{Z_0Z_1}(\beta)R_{X_0X_1}(\Theta_e)\lvert\eta\rangle_{01}.
\]
For $\alpha=\delta=\theta/2$ and $\gamma=\theta$, $\Theta_0=2\theta$ and $\Theta_1=0$; the latter branch skips its zero-angle update.
An expression that vanishes for every event record is removed during compilation.

For each residual rotation, the compiler searches backward for an equal axis, crossing commuting rotations and stopping at an anticommuting one.
A match is combined using Eq.~\eqref{eq:fused-angle}.
Activations and nonrotation instructions bound these sequences of rotations in fixed residual coordinates.
\par\endgroup

\begingroup
\setlength{\abovedisplayskip}{7pt}
\setlength{\belowdisplayskip}{7pt}
\setlength{\abovedisplayshortskip}{0pt}
\setlength{\belowdisplayshortskip}{7pt}
\begin{samepage}
\subsection{MPS layout and operator width}
After fusion, the activation and release sequence fixes $k(t)$ and $K$.
The compiler next chooses where each new active qubit is stored and which qubit is released at an active measurement.
These choices use Clifford changes of the residual representation~\cite{camps,ofd}:
\begin{equation}
\begin{aligned}
\lvert\Psi_\nu\rangle
 &=CE_\nu\lvert\psi[A_\nu]\rangle\\
 &=(CG^\dagger)(GE_\nu G^\dagger)\lvert\psi[A'_\nu]\rangle,
\qquad \lvert\psi[A'_\nu]\rangle=G\lvert\psi[A_\nu]\rangle.
\end{aligned}
\label{eq:residual-basis-change}
\end{equation}
\end{samepage}
Each later Pauli string becomes $Q_j'=GQ_jG^\dagger$.
The compiler propagates these strings through the subsequent activation and release rules and scores the MPS intervals they occupy.

The width $\Delta_j$ is the number of MPS bonds between the first and last blocks on which $Q_j$ acts nontrivially.
Group the stored positions as $(0,1)|(2,3)|(4,5)|\cdots$.
Write $Q_j=\bigotimes_r\sigma_{j,r}$ and let $\mathcal S_j$ contain the positions whose factor $\sigma_{j,r}$ is $X$, $Y$, or $Z$.
For a nonempty $\mathcal S_j$,
\begin{equation}
\Delta_j
=\left\lfloor\frac{\max\mathcal S_j}{2}\right\rfloor
 -\left\lfloor\frac{\min\mathcal S_j}{2}\right\rfloor.
\label{eq:operator-interval}
\end{equation}
If $Q_j=I$, the set $\mathcal S_j$ is empty and we set $\Delta_j=0$.
For example,
\[
\begin{array}{c|ccc}
Q_j&Z_0Z_1&Z_0Z_3&Z_0Z_5\\
\Delta_j&0&1&2
\end{array}
\]
A rotation with $\Delta=0$ updates one block; $\Delta=1$ crosses one MPS bond.
The tensor contractions on these intervals are given in Sec.~\ref{sec:mps}.

For a candidate position or pivot $g$, the score over the future window is
\begin{equation}
\begin{aligned}
\mathcal J(g)&=
 \sum_{j\in\mathcal W_{\rm rot}}\Delta_j(g)
 +2\sum_{j\in\mathcal W_{\rm meas}}\Delta_j(g),\\
g_\star&=\operatorname*{arg\,min}_g\mathcal J(g).
\end{aligned}
\label{eq:tensor-layout-cost}
\end{equation}
The window contains future rotations and active measurements.
The measurement weight two accounts for its separate probability contraction and conditional-state update.

\begin{samepage}
\emph{An activation example.---}Suppose $K=4$, with active qubits $a,b$ at stored positions $0,2$.
Positions $1,3$ are free, and a new active qubit $f$ can use either one.
Let the future window contain $R_{Z_aZ_f}(\alpha)$ followed by a measurement of $X_aX_f$.
Their spans give
\[
\begin{array}{c|cc|c}
\text{position of }f&\Delta_{\rm rot}&\Delta_{\rm meas}&\mathcal J\\\hline
1&0&0&0\\
3&1&1&3
\end{array}
\]
\end{samepage}
\begin{samepage}
The compiler chooses position $1$, placing both updates inside the first block [Fig.~\ref{fig:sm-width}].
Both candidates have the same active count $k=3$.
\end{samepage}
\begin{figure}[h]
\centering
\includegraphics[width=\linewidth]{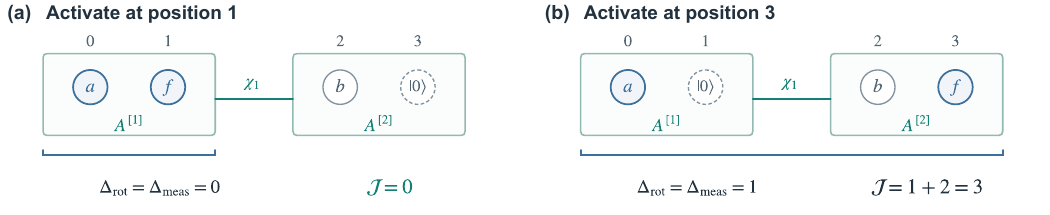}
\caption{Choosing where to activate $f$. Both panels group the positions as $(0,1)|(2,3)$, with active qubits $a,b$ at $0,2$. Blue circles mark the operands of the next rotation $Z_aZ_f$ and measurement $X_aX_f$. The bracket spans the blocks visited by both updates. (a) At position $1$, $f$ shares a block with $a$, giving $\mathcal J=0$. (b) At position $3$, both updates cross one bond, giving $\mathcal J=1+2=3$.}
\label{fig:sm-width}
\end{figure}
\par\endgroup

\begin{figure}[!h]
\centering
\includegraphics[width=\linewidth]{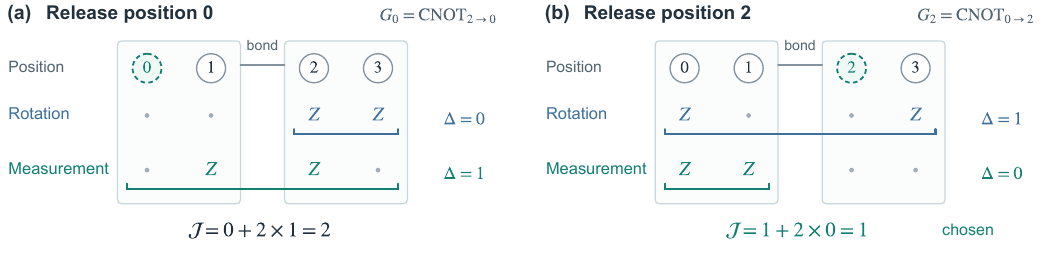}
\caption{Choosing a pivot for the measurement $Z_0Z_2$. The dashed circle marks the released position. The blue row gives the Pauli factors of the next rotation and the teal row those of the next measurement. Gray dots denote $I$. Brackets span the blocks used by that update. Releasing $0$ makes the rotation local. Releasing $2$ makes the measurement local and lowers $\mathcal J=\Delta_{\rm rot}+2\Delta_{\rm meas}$ from $2$ to $1$.}
\label{fig:sm-pivot}
\end{figure}

\emph{A measurement-pivot example.---}Now take four active positions and measure $Q=Z_0Z_2$.
Either position on which $Q$ acts can be released, using
\[
\begin{aligned}
G_0&=\mathrm{CNOT}_{2\to0},&G_0QG_0^\dagger&=Z_0,\\
G_2&=\mathrm{CNOT}_{0\to2},&G_2QG_2^\dagger&=Z_2.
\end{aligned}
\]
After the measurement, the chosen pivot is free.
Let the future window contain $R_{Z_2Z_3}(\alpha)$ followed by a measurement of $Z_1Z_2$.
For the two choices, the $Z_2$ factor in both future operators becomes
\[
Z_2\xrightarrow{\ G_0\ }Z_2,\qquad
Z_2\xrightarrow{\ G_2\ }Z_0Z_2
   \xrightarrow{\text{release }2}Z_0.
\]
The last step uses $Z_2\lvert0\rangle_2=\lvert0\rangle_2$ after release.
Figure~\ref{fig:sm-pivot} shows the resulting unsigned axes and update intervals.

Pivot $2$ gives the smaller score: its rotation crosses one bond, while its measurement becomes local to the first block.
For an incoming frame $E=I$, this choice gives $E'=X_2^m$.
The outcome $m$ supplies a factor $(-1)^m$ to the later rotation angle and measurement sign, as in Sec.~\ref{sec:active}.
Both choices release one qubit, giving $k=3$.

\emph{Compilation procedure.---}The compiler makes the position and pivot choices sequentially.
For each candidate, it copies the current Clifford map, applies the trial choice, and propagates the following Pauli strings to evaluate Eq.~\eqref{eq:tensor-layout-cost}.

Within a trial, a future activation uses the lowest free position with an active block partner, or the lowest free position otherwise.
A future measurement uses the middle of the ordered positions carrying $X$, $Y$, or $Z$, taking the upper middle when their number is even.
At each later decision, all candidates are evaluated again.
Ties keep this default choice, followed by the lowest position index.

In the pseudocode, $\mathcal I$ is the instruction list.
The candidate set $\mathcal G_j$ contains the free positions for an activation, or the positions carrying $X$, $Y$, or $Z$ for an active measurement.
$\operatorname{ApplyChoice}$ records the choice in instruction $j$ and updates the Clifford map for later instructions; $\operatorname{Encode}$ stores the position, transformed Pauli string, update interval, and angle or outcome parity.

\begin{center}
\begin{minipage}{0.96\linewidth}
\hrule\vspace{5pt}
\small
\textbf{Input:} circuit $\mathcal U$ with noise distributions and classical dependencies.\\[3pt]
\begin{tabular}{@{}r@{\quad}l@{}}
1 & $(C,E(\mathbf b),\mathcal I)\leftarrow\operatorname{Transport}(\mathcal U)$\\
2 & $\mathcal I\leftarrow\operatorname{ScheduleFuse}(\mathcal I)$\\
3 & $\mathcal I\leftarrow\operatorname{FreeQubitReduce}(\mathcal I)$\quad (Sec.~\ref{sec:active})\\
4 & $\mathcal I\leftarrow\operatorname{ResidualFuse}(\mathcal I)$\\
5 & $\mathcal P\leftarrow[\,]$\\
6 & \textbf{for each} instruction $j\in\mathcal I$ \textbf{in order:}\\
7 & \quad\textbf{if} $j$ is an activation or an active measurement:\\
8 & \qquad$g_\star\leftarrow\operatorname*{arg\,min}_{g\in\mathcal G_j}\mathcal J(g);\quad\operatorname{ApplyChoice}(j,g_\star)$\\
9 & \quad$\mathcal P.\operatorname{append}(\operatorname{Encode}(j))$\\
\end{tabular}\\[3pt]
\textbf{Output:} plan $\mathcal P$, with angle and outcome rules evaluated using $\mathbf b_\nu$.
\vspace{5pt}\hrule
\end{minipage}
\end{center}

\begin{samepage}
The searches use the following upper bounds:
\begin{center}
\begin{tabular}{@{}llr@{}}
\hline\hline
\smcell{.26}{Stage} & \smcell{.55}{Quantity} & Maximum\\\hline
\smcell{.26}{First fusion and scheduling} & \smcell{.55}{Outer iterations} & $8$\\
\smcell{.26}{} & \smcell{.55}{Rotation-scheduling sweeps per iteration} & $16$\\
\smcell{.26}{} & \smcell{.55}{Instructions per moved nonrotation block} & $64$\\\hline
\smcell{.26}{Residual fusion} & \smcell{.55}{Preceding surviving rotations examined} & $64$\\\hline
\smcell{.26}{Position and pivot search} & \smcell{.55}{Future instructions / rotations per trial} & $256$ / $256$\\
\hline\hline
\end{tabular}
\end{center}
\end{samepage}
Scheduling stops early if the sequence is unchanged; residual fusion stops at an anticommuting rotation or a fixed-coordinate sequence boundary.
The final layout and its angle rules are passed to the sampler in Sec.~\ref{sec:mps}.

\section{Block-MPS evolution and sampling}
\label{sec:mps}
The compiled plan specifies each Pauli string, rotation-angle rule, activation position, and measurement pivot.
Each shot evaluates the angles, updates the block tensors, and samples the measurement outcomes. An update crossing $\Delta_j$ MPS bonds spans $\Delta_j+1$ consecutive blocks.

\subsection{Block tensors and Pauli rotations}
The $K$ stored positions form $L=\lceil K/2\rceil$ blocks,
\begin{equation}
\lvert\phi_\nu\rangle
=\sum_{s_1,\ldots,s_L}A_\nu^{[1]s_1}A_\nu^{[2]s_2}\cdots A_\nu^{[L]s_L}
\lvert s_1\cdots s_L\rangle,
\qquad A_\nu^{[a]s_a}\in\mathbb C^{\chi_{a-1}\times\chi_a}.
\label{eq:block-state}
\end{equation}
Here $\chi_0=\chi_L=1$ and $\chi_a\leq\chi$.
Each full block has four physical labels $s_a=00,01,10,11$; the last block has two labels if $K$ is odd.
IIn a four-qubit example grouped as $(0,1)|(2,3)$, we have $s_1=u_0u_1$ and $s_2=u_2u_3$, where $u_r\in\{0,1\}$ is the bit at stored position $r$.
We suppress the shot index $\nu$ below.

If position $3$ is free, its block tensor has components $A^{[2]u_2,0}=B^{u_2}$ and $A^{[2]u_2,1}=0$.
An activation $R_{X_3}(\vartheta)$ gives
\[
\underbrace{\bigl(A^{[2]u_2,0},A^{[2]u_2,1}\bigr)}_{\text{components for qubit 3}}
=\bigl(B^{u_2},0\bigr)
\ \longrightarrow\
\bigl(\cos\vartheta\,B^{u_2},\ i\sin\vartheta\,B^{u_2}\bigr).
\]
Only the physical components change; both virtual dimensions stay the same.
This is the tensor implementation of the activation in Sec.~\ref{sec:active}.

\begin{figure}[ht]
\centering
\includegraphics[width=\linewidth]{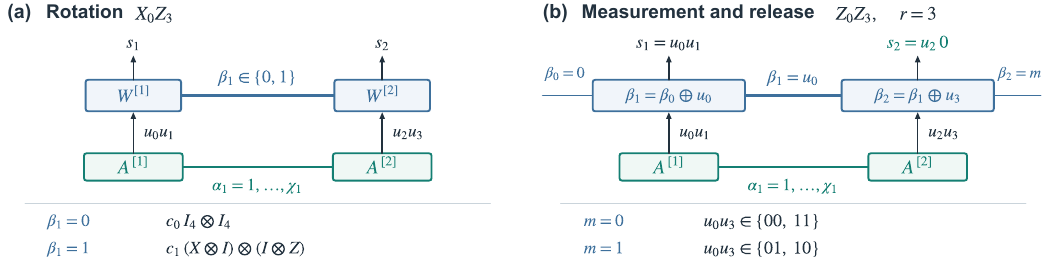}
\caption{Block-2 MPS updates on $(0,1)|(2,3)$. Physical indices run upward through the MPO. The horizontal lines carry the MPS bond $\alpha_1$ and the MPO bond $\beta_1$. (a) The $\beta_1=0$ and $\beta_1=1$ channels carry the two terms of $R_{X_0Z_3}(\vartheta)$ shown below. (b) The two XOR updates accumulate $u_0\oplus u_3$. The right boundary $\beta_2=m$ selects the indicated bit pairs, and writing $s_2=u_2\,0$ releases qubit $3$.}
\label{fig:sm-mps}
\end{figure}

\emph{A rotation as two MPO channels.---}Consider $Q=X_0Z_3$ in this block-2 example.
Its operators on the two blocks are $q^{[1]}=X\otimes I$ and $q^{[2]}=I\otimes Z$, so
\[
R_Q(\vartheta)
=\underbrace{\cos\vartheta\,I_4\otimes I_4}_{\text{identity channel}}
+\underbrace{i\sin\vartheta\,(X\otimes I)\otimes(I\otimes Z)}_{\text{Pauli term}}.
\]
On a basis state, the second channel flips $u_0$ and multiplies its amplitude by $(-1)^{u_3}$.
With $c_0=\cos\vartheta$ and $c_1=i\sin\vartheta$, write $(q^{[a]}A^{[a]})^s=\sum_u(q^{[a]})_{su}A^{[a]u}$.
Combining the MPS and MPO bonds in Fig.~\ref{fig:sm-mps}(a) gives the updated tensors
\[
\widetilde A^{[1]s_1}
=\begin{pmatrix}c_0A^{[1]s_1}&c_1(q^{[1]}A^{[1]})^{s_1}\end{pmatrix},
\qquad
\widetilde A^{[2]s_2}
=\begin{pmatrix}A^{[2]s_2}\\(q^{[2]}A^{[2]})^{s_2}\end{pmatrix}.
\]
Contracting the shared bond gives the amplitudes of $c_0\lvert\phi\rangle+c_1Q\lvert\phi\rangle$.

The same construction applies across any number of blocks. For a compiled Pauli string,
\begin{equation}
R_{Q_j}(\vartheta)=c_0I+c_1Q_j,
\qquad Q_j=\bigotimes_a q_j^{[a]}.
\label{eq:two-term-operator}
\end{equation}
Each $q_j^{[a]}$ is the tensor product of the single-qubit Pauli operators in block $a$.
Number the $\Delta_j+1$ blocks from the first to the last block on which $Q_j$ acts nontrivially as $a=1,\ldots,\Delta_j+1$.
When $\Delta_j>0$, the MPO tensors are
\begin{equation}
\begin{aligned}
W^{[1]}&=\begin{pmatrix}c_0I&c_1q_j^{[1]}\end{pmatrix},\\
W^{[a]}&=\begin{pmatrix}I&0\\0&q_j^{[a]}\end{pmatrix},
\quad 1<a<\Delta_j+1,\\
W^{[\Delta_j+1]}&=\begin{pmatrix}I\\q_j^{[\Delta_j+1]}\end{pmatrix}.
\end{aligned}
\label{eq:rotation-mpo}
\end{equation}
The virtual index $\beta_a=0,1$ selects the identity or Pauli channel throughout the interval.
Multiplying the operator-valued matrices gives exactly the two terms in Eq.~\eqref{eq:two-term-operator}.
An intermediate block with $q_j^{[a]}=I$ carries both channels onward.
Contracting each MPO tensor with its MPS tensor gives
\begin{equation}
\widetilde A^{[a]s}_{(\alpha_{a-1},\beta_{a-1}),(\alpha_a,\beta_a)}
=\sum_u(W^{[a]}_{\beta_{a-1},\beta_a})_{su}
A^{[a]u}_{\alpha_{a-1},\alpha_a}.
\label{eq:mpo-mps-contraction}
\end{equation}
Combining the MPS index $\alpha_a$ and the two-valued MPO index $\beta_a$ gives an intermediate bond dimension at most $2\chi_a$ [Fig.~\ref{fig:sm-mps}(a)].
Standard SVD compression retains a bond dimension of at most $\chi$.
For $\Delta_j=0$, the entire rotation is a local matrix update,
\[
A^{[1]s}\longleftarrow\sum_u(c_0I+c_1q_j^{[1]})_{su}A^{[1]u},
\]
with unchanged virtual dimensions.

\subsection{Measurement probabilities and conditional updates}
For the $Z_0Z_3$ measurement in Fig.~\ref{fig:sm-mps}(b), take qubits $1,2$ in $\lvert00\rangle_{12}$. In the order $(0,3)$, choose the two-qubit input
\[
\lvert\phi_{03}\rangle=\sum_{u_0,u_3=0}^{1}a_{u_0u_3}\lvert u_0u_3\rangle
=\frac{2\lvert00\rangle+\lvert01\rangle+i\lvert10\rangle+\lvert11\rangle}{\sqrt{7}}.
\]
Each basis pair has $Q=Z_0Z_3$ eigenvalue $(-1)^m$ with $m=u_0\oplus u_3$. Draw $m$ with probability \mbox{$p_m=\sum_{u_0\oplus u_3=m}|a_{u_0u_3}|^2$}. Retain the amplitudes with parity $m$ and divide them by $\sqrt{p_m}$.
\[
\begin{array}{c@{\qquad}c@{\qquad}l@{\qquad}l}
m & \text{retained }u_0u_3 & p_m & \Pi_m(Q)\lvert\phi_{03}\rangle/\sqrt{p_m}\\ \hline
0 & 00,\;11 & (4+1)/7=5/7 & (2\lvert00\rangle+\lvert11\rangle)/\sqrt{5}\\[2pt]
1 & 01,\;10 & (1+1)/7=2/7 & (\lvert01\rangle+i\lvert10\rangle)/\sqrt{2}
\end{array}
\]
The compiled pivot is $r=3$. On every retained term, $G_3=\mathrm{CNOT}_{0\to3}$ puts the same bit $m$ on this pivot. The release map in Eq.~\eqref{eq:measurement-release} acts as
\[
\lvert u_0,u_3\rangle
\xrightarrow{\ G_3\ }\lvert u_0,u_0\oplus u_3\rangle=\lvert u_0,m\rangle
\xrightarrow{\ \lvert0\rangle\langle m\rvert_3\ }\lvert u_0,0\rangle.
\]
For $m=1$, the residual output is
$(\lvert0\rangle_0+i\lvert1\rangle_0)\otimes\lvert0\rangle_3/\sqrt{2}$.
In both branches qubit $3$ factors out as $\lvert0\rangle_3$ and is free for reuse. The Pauli frame carries $X_3^m$ as in Sec.~\ref{sec:active}.

The block-MPS sampler obtains the probabilities from one expectation value. In the example, the block operators $Z\otimes I$ and $I\otimes Z$ give $\langle Z_0Z_3\rangle=p_0-p_1=3/7$. For a general active measurement, place the orthogonality center in the update interval and contract its blocks in order.
\begin{equation}
\begin{aligned}
\mathcal E_0&=I,\\
\mathcal E_a&=\sum_{s,u}(q_j^{[a]})_{su}
(A^{[a]s})^\dagger\mathcal E_{a-1}A^{[a]u},
\quad a=1,\ldots,\Delta_j+1,\\
\langle Q_j\rangle&=\operatorname{Tr}\mathcal E_{\Delta_j+1},\qquad
 p_m=\frac{1+(-1)^m\langle Q_j\rangle}{2}.
\end{aligned}
\label{eq:pauli-expectation}
\end{equation}

For the conditional update, the first box in Fig.~\ref{fig:sm-mps}(b) passes $\beta_1=u_0$ to the second. The second keeps inputs with $\beta_1\oplus u_3=m$ and writes the output $u_2\,0$.
For a general $Q_j$, first turn each nonidentity factor into $Z$ with the local basis changes
\[
b(I)=b(Z)=I,\qquad b(X)=H,\qquad b(Y)=HS^\dagger.
\]
For the rotated input label $u$ in block $a$, let $\pi_a(u)$ be the XOR of the bits at these $Z$ positions. Let $f_{a,r}(u)$ copy $u$, setting the pivot bit to zero in the block containing $r$. These two rules give the matrix elements of each MPO tensor
\begin{equation}
(W^{[a]}_{\beta_{a-1},\beta_a})_{su}
=\underbrace{\delta_{\beta_a,\,\beta_{a-1}\oplus\pi_a(u)}}_{\text{accumulate parity}}
\;\cdot\;\underbrace{\delta_{s,\,f_{a,r}(u)}}_{\text{write output}},
\qquad \beta_0=0,\quad\beta_{\Delta_j+1}=m.
\label{eq:parity-release-mpo}
\end{equation}
The right boundary selects the sampled parity $m$. Contract this MPO with the basis-transformed MPS as in Eq.~\eqref{eq:mpo-mps-contraction}, divide by $\sqrt{p_m}$, and compress the updated tensors.

\subsection{Sampling one shot}
Let $\mathcal T_{\chi,\tau}$ denote standard SVD compression of the updated bonds with maximum bond dimension $\chi$ and cutoff $\tau$, followed by normalization.
At the start of each shot, draw the Pauli faults, readout errors, and any random physical rotation signs from the noise model. At each depolarizing location, draw one Pauli operator, including the identity among the possible outcomes. At measurement $j$, the compiled parity $\ell_j$ uses bits already in the record, and $e_{M,j}$ is the readout-error bit.
\begin{center}
\begin{minipage}{0.96\linewidth}
\hrule\vspace{5pt}
\small
\textbf{Input:} plan $\mathcal P$, bond dimension $\chi$, cutoff $\tau$, and trajectory random numbers.\\[3pt]
\begin{tabular}{@{}r@{\quad}l@{}}
1 & $\lvert\phi\rangle\leftarrow\lvert0\rangle^{\otimes K};\quad
\mathbf b_\nu\leftarrow\text{sampled noise bits}$\\
2 & \textbf{for each} instruction $j\in\mathcal P$ \textbf{in order:}\\
3 & \quad\textbf{activation:} $A^{[a]}\leftarrow R_{\sigma_r}[\vartheta_j(\mathbf b_\nu)]A^{[a]}$\quad ($r$ in block $a$)\\
4 & \quad\textbf{rotation:} $\lvert\phi\rangle\leftarrow\mathcal T_{\chi,\tau}\!\left[R_{Q_j}[\vartheta_j(\mathbf b_\nu)]\lvert\phi\rangle\right]$\quad ($\vartheta_j\ne0$)\\
5 & \quad\textbf{measurement:} set $\mu_j$ to $1$ (deterministic), $0$ (free-qubit $X/Y$), or $\langle\phi|Q_j|\phi\rangle$ (active).\\
 & \qquad Draw $m\in\{0,1\}$ with probabilities $p_0=(1+\mu_j)/2$ and $p_1=(1-\mu_j)/2$.\\
6 & \qquad\textbf{if active:} $\lvert\phi\rangle\leftarrow\mathcal T_{\chi,\tau}\!\left[M_{m,r}\lvert\phi\rangle/\sqrt{p_m}\right];\quad r\leftarrow\text{free}$\\
7 & \qquad$m_{\rm phys}\leftarrow m\oplus\ell_j(\mathbf b_\nu);\quad y_j\leftarrow m_{\rm phys}\oplus e_{M,j}$.\quad Store $m$ in $\mathbf b_\nu$.\\
8 & Compute detector bits $\mathbf d$ and logical flips $\mathbf o$ from the circuit's specified parities of $\mathbf y$.\\
\end{tabular}\\[3pt]
\textbf{Output:} record $\mathbf b_\nu$ and logical-failure indicator $\mathbf 1[\operatorname{Decode}(\mathbf d)\ne\mathbf o]$.
\vspace{5pt}\hrule
\end{minipage}
\end{center}
Section~\ref{sec:protocol} gives the noise model for step~1 and the decoder used to predict the logical flips from the detector bits.

\section{Benchmark models and simulation protocol}
\label{sec:protocol}
\subsection{Noise models for the surface and bivariate bicycle codes}
We simulate two $X$-memory experiments. Figure~2 of the main paper uses a rotated surface code with coherent $ZZ$ crosstalk. At code distance $d$, we run $r=d$ rounds of syndrome extraction. Figure~3 uses the $[[72,12,6]]$ bivariate bicycle (BB) code with coherent $Y$ rotations.
\begin{center}
\begin{tabular}{@{}lrrl@{}}
\hline\hline
Circuit & Physical qubits $n$ & Peak active qubits $K$ & Coherent rotation\\ \hline
Surface code, $d=r=5$ & 49 & 17 & $R_{ZZ}(s\theta)$, $\theta=0.02$\\
Surface code, $d=r=7$ & 97 & 30 & $R_{ZZ}(s\theta)$, $\theta=0.02$\\
Surface code, $d=r=9$ & 161 & 47 & $R_{ZZ}(s\theta)$, $\theta=0.02$\\
BB code, six rounds & 144 & 45 & $R_Y(s\theta)$, $\theta=0.01$\\
\hline\hline
\end{tabular}
\end{center}
Here $s=\pm1$ is the sign of the physical rotation $R_P(s\theta)=e^{is\theta P}$. In both circuits, $p$ is the total probability of a depolarizing error on a CNOT pair.

\begin{figure}[!htb]
\centering
\includegraphics[width=\linewidth]{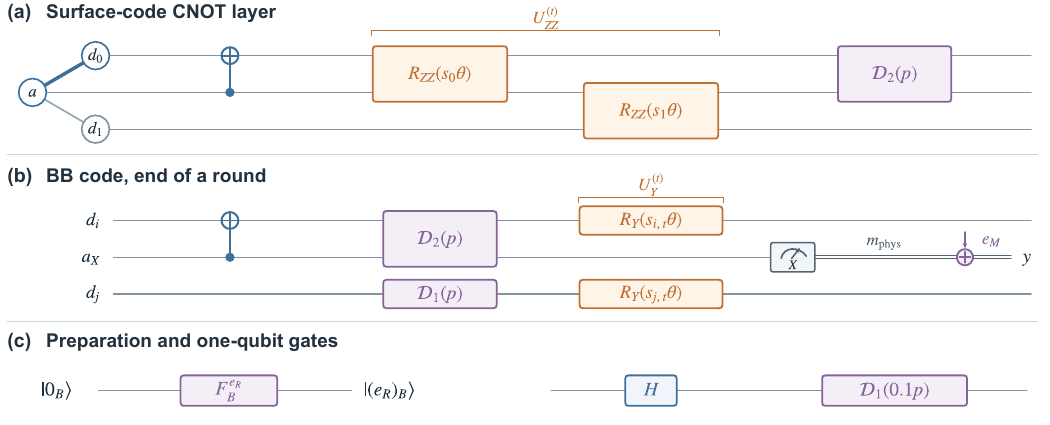}
\caption{Noise placement in the benchmark circuits. Gates act from left to right. Orange denotes coherent rotations and purple denotes stochastic errors. (a) The bold hardware edge carries the CNOT. Both edges receive a $ZZ$ rotation before $\mathcal D_2(p)$ acts on the CNOT pair. (b) At the end of a BB round, $\mathcal D_2(p)$ acts on the CNOT pair and $\mathcal D_1(p)$ on the idle data qubit. The data qubits then rotate about $Y$. Measuring the ancilla gives $m_{\rm phys}$, recorded as $y=m_{\rm phys}\oplus e_M$. (c) A preparation fault in basis $B$ and depolarization after a surface-code $H$ gate.}
\label{fig:sm-noise-circuits}
\end{figure}

\emph{Surface-code noise circuit.---}Let $\mathcal E$ contain every data--ancilla pair used by a CNOT anywhere in the ideal circuit.
After CNOT layer $t$, we apply a $ZZ$ rotation on each edge of $\mathcal E$ with at least one qubit that took part in a CNOT in that layer. The set of these edges and the product of their rotations are
\[
\begin{aligned}
\mathcal E_t&=\{(u,v)\in\mathcal E:\ u\text{ or }v\text{ participates in layer }t\},\\
U_{ZZ}^{(t)}&=\prod_{(u,v)\in\mathcal E_t}R_{Z_uZ_v}(s_{uv,t}\theta).
\end{aligned}
\]
After $U_{ZZ}^{(t)}$, we apply the two-qubit depolarizing channel $\mathcal D_2(p)$ to each CNOT pair. In Fig.~\ref{fig:sm-noise-circuits}(a), $a$ controls a CNOT on $d_0$. Both $(a,d_0)$ and $(a,d_1)$ receive a $ZZ$ rotation because both edges meet $a$. We then apply $\mathcal D_2(p)$ to the CNOT pair $(a,d_0)$.

\emph{BB noise circuit.---}Each CNOT pair receives $\mathcal D_2(p)$ after the gate. A qubit scheduled to idle receives $\mathcal D_1(p)$. After the last CNOT layer of round $t$ and its Pauli errors, we apply a $Y$ rotation to each of the $72$ data qubits,
\[
U_Y^{(t)}=\prod_{q=0}^{71}R_{Y_q}(s_{q,t}\theta),\qquad \theta=0.01.
\]
We then measure the $X$-check ancillas in the $X$ basis [Fig.~\ref{fig:sm-noise-circuits}(b)]. The six rounds contain $72\times6=432$ physical $Y$ rotations.

\emph{Pauli errors, preparation, and readout.---}At each depolarizing location we sample one Pauli operator. For total error probability $q$, the identity has probability $1-q$. On one qubit, $X$, $Y$, and $Z$ each have probability $q/3$. On a qubit pair, each of the $15$ Pauli operators other than $II$ has probability $q/15$.
Different noise locations are sampled independently.
The surface-code rates follow Table~I of Ref.~\cite{harper}, and the BB rates follow Ref.~\cite{bb}.
\begin{center}
\begin{tabular}{@{}lcc@{}}
\hline\hline
Location & Surface code & BB code\\ \hline
After a one-qubit Clifford gate & $\mathcal D_1(0.1p)$ & ---\\
Scheduled idle & --- & $\mathcal D_1(p)$\\
CNOT pair & $\mathcal D_2(p)$ after the $ZZ$ layer & $\mathcal D_2(p)$ after the gate\\
Preparation or reset, $p_R$ & $2p$ & $p$\\
Readout-bit flip, $p_M$ & $5p$ & $p$\\
\hline\hline
\end{tabular}
\end{center}
For example, at $p=0.005$ the surface-code channels are $\mathcal D_1(0.0005)$ and $\mathcal D_2(0.005)$, with $p_R=0.01$ and $p_M=0.025$.

We sample the preparation and readout error bits $e_R$ and $e_M$ independently [Fig.~\ref{fig:sm-noise-circuits}(b,c)]. Each bit is 1 when its error occurs and 0 otherwise. For $B=Z$ or $X$,
\[
\begin{aligned}
\lvert0_B\rangle&\longmapsto F_B^{e_R}\lvert0_B\rangle,
&&\Pr(e_R=1)=p_R,\\
y&=m_{\rm phys}\oplus e_M,
&&\Pr(e_M=1)=p_M.
\end{aligned}
\]
A preparation error applies $F_Z=X$ to $\lvert0_Z\rangle=\lvert0\rangle$ or $F_X=Z$ to $\lvert0_X\rangle=\lvert+\rangle$. At readout, we record the physical outcome $m_{\rm phys}$ when $e_M=0$ and flip its bit when $e_M=1$.

\subsection{Rotation signs and reference simulations}
Figures~1(d,e), 4, and~6, as well as the fixed-sign curves in Fig.~5, use the same surface-code circuit with $d=r=5$, $p=0.005$, and $\theta=0.02$. Every $ZZ$ rotation has sign $s_j=+1$. We estimate the reference logical error rate by sampling this circuit with an untruncated residual state vector.

For Figs.~2 and~3, each shot draws an independent sign $s_j=\pm1$ with equal probabilities for every physical rotation. CAMPS uses these signed angles in Eq.~\eqref{eq:fused-angle}. For example, two consecutive rotations about the same $P$ fuse to an angle $(s_1+s_2)\theta$. This angle is zero with probability $1/2$ and takes each of the values $+2\theta$ and $-2\theta$ with probability $1/4$. Averaging one rotation over its two signs gives
\begin{equation}
\frac12\sum_{s=\pm1}R_P(s\theta)\rho R_P(s\theta)^\dagger
=\cos^2\theta\,\rho+\sin^2\theta\,P\rho P.
\label{eq:sign-average}
\end{equation}
The probability of applying $P$ is $\sin^2\theta$, about $4.00\times10^{-4}$ for the surface code and $1.00\times10^{-4}$ for BB. Since the signs are independent, replacing each rotation by this channel gives the circuit averaged over all signs. We use a stabilizer simulator to sample that circuit and estimate the reference logical error rate labeled PTA (Pauli-twirling approximation) in Figs.~1(d), 2, 3, and~5

\subsection{Decoding and logical error rates}
For each circuit and value of $p$, all simulators use the same decoder. We build it from the detector error model of the Pauli-twirled circuit. This model lists each error, its probability, and the detector and logical bits it flips.

In each shot, the circuit specifies which measurement bits in $\mathbf y_\nu$ (Sec.~\ref{sec:mps}) to add modulo two. These sums give the detector outcomes $\mathbf d_\nu$ and the observed logical flips $\mathbf o_\nu$. The decoder predicts the logical flips from $\mathbf d_\nu$. A shot fails if its prediction differs from $\mathbf o_\nu$. The estimated logical error rate is the number of failures divided by the number of shots $N$.
\[
\widehat{\mathbf o}_\nu=\operatorname{Decode}(\mathbf d_\nu),\qquad
L_\nu=\mathbf 1[\widehat{\mathbf o}_\nu\ne\mathbf o_\nu],\qquad
\widehat p_L=\frac1N\sum_{\nu=1}^{N}L_\nu .
\]
The surface-code experiment compares one logical $X$ bit. In the BB experiment, one or more wrong predictions among the $12$ logical $X$ bits after six rounds count as one block failure.
\begin{center}
\small
\begin{tabular}{@{}lll@{}}
\hline\hline
\smcell{.12}{Code} & \smcell{.34}{Detector error model} & \smcell{.47}{Decoder}\\ \hline
\smcell{.12}{Surface} & \smcell{.34}{Decomposed into matching edges} & \smcell{.47}{PyMatching~\cite{pymatching} minimum-weight perfect matching, with correlated matching disabled}\\
\smcell{.12}{BB} & \smcell{.34}{Undecomposed (one column per error)} & \smcell{.47}{Belief propagation (up to $50$ iterations) with ordered-statistics decoding (order $4$)~\cite{bposd}}\\
\hline\hline
\end{tabular}
\end{center}

\emph{Shots per data point.---}Compiled and uncompiled CAMPS evolve the Clifford-augmented MPS with and without compile-time reduction, respectively. Physical MPS denotes an MPS on the physical circuit qubits.
\begin{center}
\small
\begin{tabular}{@{}llll@{}}
\hline\hline
\smcell{.28}{Comparison} & \smcell{.17}{$p$} & \smcell{.32}{Shots per tensor simulation} & \smcell{.16}{Reference shots}\\ \hline
\smcell{.28}{Fig.~1(d)} & \smcell{.17}{$0.005$} & \smcell{.32}{$5\times10^5$ (compiled and uncompiled CAMPS); $10^4$ (physical MPS)} & \smcell{.16}{$5\times10^5$ each (state vector, PTA)}\\
\smcell{.28}{Fig.~2, $\chi=8$} & \smcell{.17}{$0.001,0.002,0.003$} & \smcell{.32}{$10^5$} & \smcell{.16}{$10^6$ (PTA)}\\
 & \smcell{.17}{$>0.003$} & \smcell{.32}{$10^4$} & \smcell{.16}{$10^6$ (PTA)}\\
\smcell{.28}{Fig.~3, $\chi=4$} & \smcell{.17}{$0.001$} & \smcell{.32}{$2\times10^5$} & \smcell{.16}{$10^6$ (PTA)}\\
 & \smcell{.17}{$>0.001$} & \smcell{.32}{$10^4$} & \smcell{.16}{$10^6$ (PTA)}\\
\smcell{.28}{Fig.~4, compiler comparison} & \smcell{.17}{$0.005$} & \smcell{.32}{$10^6$ per compiler at each $\chi$} & \smcell{.16}{$10^6$ (state vector)}\\
\smcell{.28}{Fig.~5, fixed signs} & \smcell{.17}{$0.005$} & \smcell{.32}{$5\times10^5$ per method at each $\chi$} & \smcell{.16}{$5\times10^5$ (state vector)}\\
\smcell{.28}{Fig.~5, random signs} & \smcell{.17}{$0.005$} & \smcell{.32}{$5\times10^5$ per method at each $\chi$} & \smcell{.16}{$5\times10^5$ (PTA)}\\
\smcell{.28}{Fig.~6, block-size comparison} & \smcell{.17}{$0.005$} & \smcell{.32}{$10^6$ per block size at each $\chi$} & \smcell{.16}{$10^6$ (state vector)}\\
\hline\hline
\end{tabular}
\end{center}
\bibliography{references}

%% file: end_matter.tex
\setlength{\textfloatsep}{10pt plus 2pt minus 2pt}
\setlength{\intextsep}{8pt plus 2pt minus 2pt}
\setlength{\abovecaptionskip}{6pt}
\setlength{\belowcaptionskip}{0pt}
\emph{Compiler comparison.---}Different Clifford representations of the same physical state can require different MPS bond dimensions and update costs.
On the fixed-sign $ZZ$ surface-code benchmark of Fig.~\ref{fig:camps}(d), with $d=r=5$, $\theta=0.02$, and $p=0.005$, we compare the representations $CE(\mathbf b)$ from compiled CAMPS and Clifft~\cite{clifft} using the same block-2 MPS backend.
Both compilations reduce the $49$ physical qubits to a peak of $K=17$ active qubits.
Beyond qubit recycling, compiled CAMPS selects activation positions and measurement pivots using subsequent operator widths and combines rotations through two fusion stages.

\begin{figure}[h]
\centering
\includegraphics[width=\columnwidth]{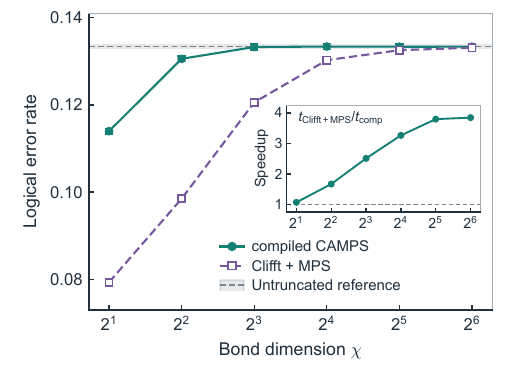}
\caption{Logical error rate $p_L$ versus bond dimension $\chi$ for compiled CAMPS and Clifft compilation + MPS on the same block-2 MPS backend ($K=17$) in the benchmark of Fig.~\ref{fig:camps}(d).
Bars and shading show 95\% confidence intervals ($10^6$ shots per point).
Inset: median sampling speedup $t_{\mathrm{Clifft+MPS}}/t_{\mathrm{comp}}$.}
\label{fig:compiler-accuracy}
\end{figure}

Figure~\ref{fig:compiler-accuracy} shows that compiled CAMPS agrees with the untruncated reference within sampling uncertainty at $\chi=8$, and its LER $p_L$ remains stable as $\chi$ increases.
At $\chi=8$, Clifft compilation followed by MPS evolution underestimates $p_L$ by approximately $10\%$ and approaches the reference more gradually at larger $\chi$.
Compiled CAMPS thus yields better convergence with $\chi$ and, as shown in the inset, faster sampling.

\begin{figure}[!t]
\centering
\includegraphics[width=\columnwidth]{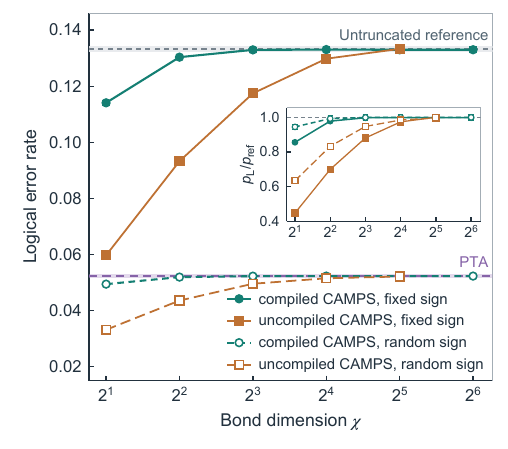}
\caption{Logical error rate $p_L$ versus bond dimension $\chi$ for fixed- and random-sign $ZZ$ rotations in the surface-code circuit of Fig.~\ref{fig:camps}(d).
Gray and purple dashed lines mark the untruncated $K=17$ reference for fixed signs and the exact PTA reference for random signs.
Inset: normalized ratio $p_L/p_{\mathrm{ref}}$ relative to each reference.
Bars and shading show 95\% confidence intervals.}
\label{fig:sign-convergence}
\end{figure}

\emph{Fixed- and random-sign convergence.---}Figure~\ref{fig:sign-convergence} compares convergence with bond dimension $\chi$ for fixed- and random-sign coherent $ZZ$ rotations in rotated surface-code $X$ memory at $d=r=5$, $p=0.005$, and $\theta=0.02$.
Fixed-sign rotations raise the untruncated $K=17$ state-vector LER to more than $2.5$ times the random-sign PTA value.
The inset plots the normalized ratio $p_L/p_{\mathrm{ref}}$ relative to the state-vector reference for fixed signs and the PTA reference for random signs.
Fixed signs give a lower ratio at $\chi=2$, but both compiled CAMPS curves are close to unity at $\chi=4$ and converge within sampling uncertainty at $\chi=8$, whereas uncompiled CAMPS requires $\chi=32$ in both settings.

\begin{figure}[!t]
\centering
\includegraphics[width=\columnwidth]{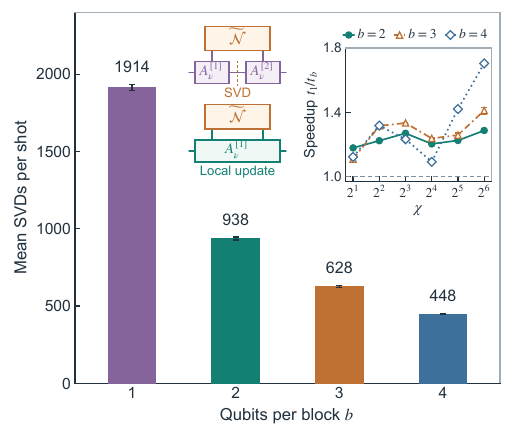}
\caption{Mean number of SVDs per shot versus block size $b$ at $\chi=8$ in the surface-code circuit of Fig.~\ref{fig:camps}(d).
Left inset: grouping qubits turns a two-qubit residual rotation into a local tensor update.
Right inset: median sampling speedup $t_1/t_b$ versus bond dimension $\chi$, where $t_b$ is the sampling time with $b$ qubits per block.
Error bars show 95\% confidence intervals.}
\label{fig:block-mps}
\end{figure}

\emph{Block MPS.---}Grouping neighboring qubits into one tensor further reduces the cost of evolving the residual state.
In a single-qubit MPS, a rotation coupling two neighboring qubits requires splitting the updated tensor by a singular-value decomposition (SVD).
Placing both qubits in the same block turns this step into a local tensor update, as illustrated in the left inset of Fig.~\ref{fig:block-mps}.
More generally, a block of $b$ qubits has a physical index of dimension $2^b$, with $\chi$ limiting only the bonds between blocks.
Rotations and measurements supported within one block act directly on this physical index, avoiding SVDs across internal qubit boundaries.

Figure~\ref{fig:block-mps} compares $b=1$, $2$, $3$, and $4$ on the fixed-sign $ZZ$ benchmark of Fig.~\ref{fig:camps}(d), optimizing the compiled basis for each block size.
Block-2 roughly halves the mean number of SVDs per shot, from $1914$ to $938$, and $b=3$ and $4$ reduce it further to $628$ and $448$.
In the right inset, blocking speeds up sampling relative to $b=1$ across all $\chi$.
Because larger blocks reduce SVDs at the cost of a $2^b$ physical dimension, the fastest grouping depends on circuit structure and $\chi$.
All other benchmarks here use block-2.
The block partition can also be optimized together with the Clifford representation, for example by varying block sizes along the MPS or as the circuit evolves.